\PassOptionsToPackage{colorlinks=true,citecolor=blue,linkcolor=blue,urlcolor=blue}{hyperref}

\documentclass{article}

\usepackage{arxiv}

\usepackage[utf8]{inputenc} 
\usepackage[T1]{fontenc}    
\usepackage{hyperref}       
\usepackage{url}            
\usepackage{booktabs}       
\usepackage{amsfonts}       
\usepackage{nicefrac}       
\usepackage{microtype}      
\usepackage{lipsum}		
\usepackage{graphicx}
\usepackage{doi}

\usepackage{array}
\usepackage[table]{xcolor}
\definecolor{highlight}{HTML}{D1D1D1}

\usepackage{natbib}

\newcommand{\R}{\mathbb{R}}

\usepackage{esvect}
\usepackage{amsmath}
\usepackage{subcaption}

\title{Bayesian Non-Negative Matrix Factorization with Correlated Mutation Type Probabilities for Mutational Signatures}

\author{%
  {\hypersetup{urlcolor=black}%
   \href{https://orcid.org/0009-0008-8944-0526}{%
     \textcolor{black}{\includegraphics[scale=0.06]{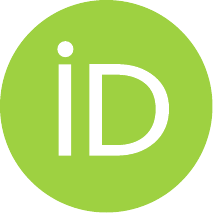}\hspace{1mm}Iris Y.~Lang}}}%
  \\[1ex]
  Harvard University\\
  \texttt{ilang@college.harvard.edu}
  \And
  {\hypersetup{urlcolor=black}%
   \href{https://orcid.org/0000-0002-1411-8750}{%
     \textcolor{black}{\includegraphics[scale=0.06]{orcid.pdf}\hspace{1mm}Jenna M.~Landy}}}%
  \\[1ex]
  Harvard University\\
  \texttt{jlandy@g.harvard.edu}
  \And
  {\hypersetup{urlcolor=black}%
   \href{https://orcid.org/0000-0002-8783-5961}{%
     \textcolor{black}{\includegraphics[scale=0.06]{orcid.pdf}\hspace{1mm}Giovanni Parmigiani}}}%
  \\[1ex]
  Dana–Farber Cancer Institute\\
  Harvard University\\
  \texttt{gp@jimmy.harvard.edu}
}

\date{}

\renewcommand{\headeright}{}
\renewcommand{\undertitle}{}

\renewcommand{\shorttitle}{corBayesNMF}

\hypersetup{
pdftitle={Bayesian Non-Negative Matrix Factorization with Correlated Mutation Type Probabilities for Mutational Signatures},
pdfsubject={q-bio.NC, q-bio.QM},
pdfauthor={Iris Y.~Lang, Jenna M. ~Landy, Giovanani Parmigiani},
pdfkeywords={Non-negative matrix factorization, Gibbs sampling, Mutational signatures
analysis},
}

\begin{document}
\maketitle

\begin{abstract}
    Somatic mutations, or alterations in DNA of a somatic cell, are key markers of cancer. In recent years, mutational signature analysis has become a prominent field of study within cancer research, commonly with Nonnegative Matrix Factorization (NMF) and Bayesian NMF. However, current methods assume independence across mutation types in the signatures matrix. This paper expands upon current Bayesian NMF methodologies by proposing novel methods that account for the dependencies between the mutation types. First, we implement the Bayesian NMF specification with a Multivariate Truncated Normal prior on the signatures matrix in order to model the covariance structure using external information, in our case estimated from the COSMIC signatures database. This model converges in fewer iterations, using MCMC, when compared to a model with independent Truncated Normal priors on elements of the signatures matrix and results in improvements in accuracy, especially on small sample sizes. In addition, we develop a hierarchical model that allows the covariance structure of the signatures matrix to be discovered rather than specified upfront, giving the algorithm more flexibility. This flexibility for the algorithm to learn the dependence structure of the signatures allows a better understanding of biological interactions and how these change across different types of cancer. The code for this project is contributed to an open-source R software package. Our work lays the groundwork for future research to incorporate dependency structure across mutation types in the signatures matrix and is also applicable to any use of NMF beyond just single-base substitution (SBS) mutational signatures.
\end{abstract}

\keywords{Non-negative matrix factorization \and Gibbs sampling \and Mutational signatures
analysis \and Multivariate truncated normal sampling}

\section{Introduction}
\label{sec:Introduction}

\subsection{Mutational Signatures} 
\label{sec:Mutational Signatures}


Somatic mutations, or alterations to DNA of a somatic cell, are critical markers of cancer and are integral to understanding its development and progression. These mutations arise from mutational processes, which are biological mechanisms that cause changes to the DNA sequence over time. Mutational processes can be endogenous (e.g., DNA replication errors, reactive oxygen species) or exogenous (e.g., UV radiation exposure, tobacco smoke, chemicals) \citep{martincorena2015somatic}. Each mutational process leaves a characteristic imprint, known as a mutational signature, on the cancer genome \citep{helleday2014mechanisms}. 

Broadly, somatic mutations can be categorized into two types: driver mutations and passenger mutations. Driver mutations are considered to be the key genetic alterations that directly contribute to cancer development by conferring a selective advantage to cancer cells, allowing them to proliferate. Passenger mutations, on the other hand, are byproducts of mutational processes, occurring alongside driver mutations and preserving rich historical information about the mutational processes that have acted on the genome. In the past, cancer research primarily focused on driver mutations \citep{helleday2014mechanisms, nik-zainal2012mutational}. When only looking at driver mutations, the mutational patterns captured are influenced by selective pressures, which can obscure the underlying patterns generated by DNA damage and repair processes \citep{nik-zainal2012mutational}. However, of the thousands of mutations that make up cancer genomes, the vast majority of them are passenger mutations that arose either prior to or shortly after the emergence of a driver mutation \citep{van-hoeck2019portrait}. This abundance of passenger mutations makes them particularly valuable for studying mutational processes because they provide a large dataset to infer patterns and mechanisms without the selective pressures that affect driver mutations \citep{martincorena2015somatic}.

In recent years, mutational signature analysis has become a prominent field of study within cancer research that focuses on uncovering such patterns from passenger somatic mutations data. Mutational signatures are typically represented as probability distributions over mutation types. Though signatures have been analyzed across many classes of mutations, this work focuses on single base substitutions (SBS) mutations with trinucleotide context, which are typically categorized into $96$ types. There are six possible single base substitutions (C>A, C>G, C>T, T>A, T>C, T>G), where base pairing rules convert any purine mutation (such as G> or A>) to its corresponding pyrimidine-based representation. Additionally, looking at the four possible bases immediately 5' and 3' of the mutated site gives $4 \times 4 = 16$ possible combinations for the trinucleotide context \citep{alexandrov2013deciphering}. This results in the alphabet of the aforementioned total of $4 \times 4 \times 6 = 96$ mutation types.

With the advent of genome sequencing technologies and large-scale cancer genome projects, such as The Cancer Genome Atlas (TCGA) and the Pan-Cancer Analysis of Whole Genomes (PCAWG), researchers have been able to catalog mutational signatures across a wide variety of cancers \citep{alexandrov2020repertoire, TCGAPanCancer2013}. These efforts have identified dozens of distinct mutational signatures which may be associated with specific exogenous exposures or endogenous processes. For example, UV-induced mutational processes generate a signature dominated by C>T substitutions at dipyrimdine sites \citep{pfeifer2005mutations} while smoking-related processes produce G>T transversions \citep{pfeifer2010environmental}. However, not all mutations have been linked to specific biological processes, and more research is required. Nonetheless, the applications of mutational signature analysis to cancer diagnosis will have tremendous clinical impact \citep{van-hoeck2019portrait}.

\subsection{NMF for Mutational Signatures}

Mutational signature analysis models a tumor's mutational landscape as the combined effect of multiple mutational processes working simultaneously. These processes are represented as latent signatures that must be computationally discovered, generally through unsupervised learning and matrix decompositions. Currently, there is not yet an agreed upon universal set of mutational signatures, however, there are many existing methods that model mutational signatures \citep{alexandrov2013deciphering}. This computational discovery of mutational signatures from mutation count data, without relying on prior knowledge or predefined references, is known as de novo extraction. De novo extraction is particularly valuable for identifying novel signatures and uncovering unknown mutational processes in cancer datasets.

Non-Negative Matrix Factorization (NMF) is a commonly used technique to model mutational signatures. Suppose a mutations count matrix $M$ has $K$ mutation types and $G$ tumor samples, and suppose there are $N$ underlying mutational signatures. The idea behind using NMF is to decompose the mutation counts matrix ($M \in \R_{\geq 0}^{K \times G}$ matrix which contains mutation counts for tumor genomes) into the product of a signatures matrix ($P \in \R_{\geq 0}^{K \times N}$ matrix which relates mutation types with underlying, latent mutational signatures) with an exposures matrix ($E \in \R_{\geq 0}^{N \times G}$ matrix which relates mutational signatures with tumor genomes). That is, NMF will estimate the signatures and exposures matrices so as to reconstruct $M \approx PE$. 

Each of the $N$ columns in the signatures ($P$) matrix can be thought of as a probability mass function over the $K$ mutation types for a given signature. That is, the n-th column of $P$ (of length $K$) would be the probability mass function over the $K$ mutation types for the n-th mutational signature \citep{alexandrov2013deciphering}. Typically, $K = 96$, using the alphabet defined in the previous section. An element $E_{ng}$ in the exposures matrix represents the contribution of mutational signature $n$ to tumor genome $g$, that is, the ``exposure" of tumor genome $g$ to mutational signature $n$. As such, this formulation reflects the additive nature of mutational signatures where the observed mutation counts in a sample are a linear combination of mutational signatures. This mathematical formulation parallels the biological perspective of a tumor's mutational landscape being a weighted combination of contributions from signatures and intensities of exposures from multiple underlying processes, throughout the entire lineage of cells leading up to the cancer cell.

Using the standard NMF procedure, $P$ and $E$ can be estimated via an iterative algorithm that seeks to minimize some norm (typically, the Kullback-Leibler (KL)-divergence or the Frobenius norm) relating to the difference between $M$ and $PE$. For example, Lee and Seung's multiplicative gradient descent update rules ensure convergence to a local minimum of the objective function \citep{lee2001algorithms}:
$$P_{ik} \leftarrow P_{ik}\frac{(ME^T)_{ik}}{(PEE^T)_{ik}}, \;\;\; E_{ik} \leftarrow E_{kj}\frac{(P^TM)_{kj}}{(P^TPE)_{kj}}$$

These updates are guaranteed to maintain non-negativity and converge to a local minimum of the cost function under certain conditions. While convergence to a global minimum cannot be guaranteed due to the non-convexity of the optimization problem and the fact that finding an exact solution to NMF is NP-hard, empirical studies have shown that the algorithm often yields meaningful and interpretable results in practice \citep{vavasis2007complexityOfNMF}. As such, NMF is typically solved using iterative optimization techniques despite the fact that a global minimum is not guaranteed.

\subsection{Initial Applications of NMF to Mutational Signatures}

One of the first significant applications of mutational signature analysis using NMF was by Nik-Zainal et al. (2012), who generated somatic mutation catalogs from 21 breast cancers. They applied NMF to these datasets to decompose the mutation counts into five distinct mutational signatures, each corresponding to specific biological processes, such as aging or defective DNA repair mechanisms \citep{nik-zainal2012mutational}. This groundbreaking study demonstrated the power of NMF in revealing the latent mutational processes shaping the cancer genome, paving the way for broader applications of this approach in cancer research.

Building on this work, Alexandrov et al. (2013) extended the application of NMF to 30 different cancer types, encompassing more than 7,000 samples, to perform a comprehensive analysis of mutational processes across a wide range of tumors. Using NMF, they identified 21 distinct mutational signatures, each characterized by unique patterns of single base substitutions (SBS) and their trinucleotide contexts \citep{alexandrov2013deciphering}. These signatures were linked to specific mutational processes, such as UV light exposure (high frequency of C>T transitions at dipyrimidine sites) and tobacco smoking (many G>T transversions due to bulky DNA adducts). Other signatures reflected endogenous processes, such as the spontaneous deamination of methylated cytosines or the activity of APOBEC enzymes. Alexandrov et al.’s study demonstrated that multiple mutational processes could act concurrently within a single tumor, with varying levels of contribution, emphasizing the complexity of the mutational landscape \citep{alexandrov2013signatures}.

These studies established NMF as a foundational technique for mutational signature analysis and highlighted its utility in uncovering both dominant and subtle mutational processes. Furthermore, the systematic approach adopted by Alexandrov et al. (2013) \citep{alexandrov2013deciphering} led to the creation of the Catalogue Of Somatic Mutations In Cancer (COSMIC) mutational signature database by Tate et al. (2019) \citep{tate2019cosmic}, which remains a critical resource for researchers studying mutational processes and their implications in cancer biology.

Aside from the basic NMF described above, there are numerous other variants of NMF, some of which we will discuss in the following sections.

\subsection{Probabilistic Interpretation}

The matrix factorization of the $M$ matrix in NMF can also be framed as a probabilistic model. In this framework, the observed mutation counts in $M$ are assumed to follow a probabilistic distribution whose parameters are determined by the latent variables in $P$ and $E$. That is, we can think of the elements of $P$ and $E$ as parameters that we seek to estimate.

It is natural to use a Poisson distribution to model the mutation counts matrix:
$$M_{ij} \sim \text{Poisson}((PE)_{ij})$$

where $(PE)_{ij} = \sum_{k=1}^N P_{ik}E_{kj}$ represents the weighted contribution of all latent factors (mutational signatures) to the observed count. Note that now, we have:
$$\mathbb{E}[M_{ij}] = (PE)_{ij}$$

To fit the Poisson NMF model, we want to to maximize the likelihood of the observed data matrix, $M$. This is equivalent to minimizing the negative log-likelihood:
$$L(P, E) = -\sum_{i=1}^K \sum_{j=1}^G [M_{ij} \log((PE)_{ij}) - (PE)_{ij}],$$
$$P \geq 0, \; \; E \geq 0$$

It has been shown that minimizing the Kullback-Leibler (KL) Divergence between the observed matrix $M$ and the product $PE$ via gradient descent is equivalent to maximizing the likelihood:
$$M_{kg} \sim \text{Poisson}((PE)_{kg})$$

using the Expectation-Maximization (EM) algorithm \citep{cemgil2009probabilistic, landy2025bayesnmf, berman2015nonnegative}. That is:
$$\arg\min_{P,E} KL(M \parallel PE) = \arg\max_{P, E} \sum_{k,g} p(M_{kg} | \lambda = (PE)_{kg})$$

which illustrates that such a probabilistic representation using a Poisson likelihood has an equivalence with standard NMF.

Similarly, under the normal approximation:
$$M_{kg} \sim \text{Normal}((PE)_{kg}, \sigma^2)$$

Schmidt et al. (2009) showed that minimizing the Frobenius norm, also known as the Mean Squared Error (MSE), is equivalent to maximizing the Normal likelihood of $M$ centered at $PE$. That is:
$$\arg\min_{P,E} \|M - PE\|_F^2 = \arg\max_{P, E} \sum_{k,g} p(M_{kg} | \mu = (PE)_{kg}, \sigma^2)$$

illustrating that such a probabilistic representation using a Normal likelihood has an equivalence with standard NMF.

\subsection{Extensions to Generalized Models}

There are instances in which a Poisson model may not be the best choice of model. For example, when we encounter overdispersion in the data, that is, when the variance exceeds the mean (since the mean and variance are equal in a Poisson model), an alternative model may be required. Therefore, there have been other probabilistic NMF extensions developed that use different likelihood functions to account for such issues. 

The Negative Binomial distribution is one extension to the Poisson model in probabilistic NMF, incorporating an additional dispersion parameter that makes it more suitable for overdispersed count data \citep{pelizzola2023model}. The Negative Binomial can be derived as a Gamma-Poisson mixture, allowing it to capture extra-Poisson variability often seen in count data.

For datasets where counts are not strictly required or where computational simplicity is prioritized, the Gaussian distribution is often used as an alternative. While less natural for count data, the Gaussian model is well-suited for continuous datasets or as an approximation that simplifies calculations \citep{schmidt2009bayesian}. We will discuss the Normal approximation for the Poisson model more in later sections.

Another important extension is the use of zero-inflated models. Many biological datasets, particularly those that are sparse, have an excess of zeros that models like the Poisson or Negative Binomial distributions cannot adequately represent \citep{abe2017nonnegative}. Zero-inflated models address this by introducing a parameter that models the probability of an excess zero occurrence, distinguishing between structural zeros and zeros that arise from the underlying count process \citep{favero2021zero}. As such, these models result in a more accurate and nuanced representation of the data and are especially useful in contexts with sparse data.

\subsection{Bayesian NMF}

From the probabilistic model, we can extend it to a Bayesian NMF framework by imposing priors on the elements of the $P$ (signatures) and $E$ (exposures) matrices. This extension enables researchers to account for prior knowledge or impose constraints, such as sparsity or hierarchical structures, while providing uncertainty quantification (such as credible intervals) for the estimated parameters. Bayesian NMF can be solved using Markov Chain Monte Carlo (MCMC) methods, such as Gibbs samplers \citep{schmidt2009bayesian}. Gibbs sampling is a widely used MCMC method for sampling from complex, high-dimensional posterior distributions. The core idea is to iteratively sample each parameter (or group of parameters) from its full conditional distribution (the distribution of that parameter given the current values of all other parameters and the observed data). It can be shown, theoretically, that the sequence of samples produced by a Gibbs sampler converges to samples drawn from the true joint posterior distribution \citep{Robert1999gibbs}.

The current standard for Bayesian NMF applied to mutational signature analysis employs a Poisson likelihood, often paired with Gamma priors on $P$ and $E$ to encourage non-negativity and account for overdispersion in mutation counts \citep{alexandrov2013deciphering, pelizzola2023model}. In some cases, researchers have explored Dirichlet priors, which impose additional constraints on the proportions of contributions across samples or signatures \citep{schmidt2009bayesian}. Hierarchical extensions, such as the Hierarchical Dirichlet Process Mixture Models, have also been used to allow for greater flexibility in identifying latent factors, particularly in datasets with unknown or varying numbers of signatures \citep{blei2012latent}.

However, a Gibbs sampler for a Bayesian model with a Poisson likelihood requires a computationally intensive Poisson augmentation. Recently, \citet{landy2025bayesnmf} showed that for a fixed rank, models built upon the Poisson and Normal likelihoods reach very similar solutions. This work utilizes the Normal likelihood to reduce the overall computational complexity of our proposed model.

Our work builds off of the Normal - Truncated Normal model as implemented in the \texttt{bayesNMF} R package. The likelihood of the $M$ matrix is modeled as Gaussian, with an Inverse Gamma prior $\sigma^2_k$ (corresponding to mutation type $k$) imposed on the variance of each element in row $k$ of matrix $M$:
$$M_{kg} \sim \text{Normal}((PE)_{kg}, \sigma_k^2), \; \; \sigma_k^2 \sim \text{InvGamma}(\alpha_k, \beta_k)$$

The priors on $P$ and $E$ are modeled using Truncated Normal distributions, truncated between zero and infinity:
$$P_{kn} \sim \text{TruncNorm}(\mu_{kn}^P, \sigma_{kn}^{2P}, 0, \infty)$$
$$E_{ng} \sim \text{TruncNorm}(\mu_{ng}^E, \sigma_{ng}^{2E}, 0, \infty)$$

For the Gibbs updates of $P$ and $E$, their full conditional distributions are again Truncated Normal distributions (see Appendix~\ref{Appendix: Norm-Trunc Norm Posterior Derivation}). The posterior on $\sigma_k^2$ remains an Inverse Gamma distribution. Prior parameters $\mu^P_{kn}, \sigma^{2P}_{kn}, \mu^E_{ng}, \sigma^{2E}_{ng}$ each follow hyperprior distributions (see \citet{landy2025bayesnmf} for details).

\subsection{Software for Bayesian NMF}

In this section, we provide an overview of existing software for Bayesian NMF, particularly with application to mutational signature analysis. One of the most well-known computational packages is \texttt{SigProfiler} which is the Wellcome Sanger Institute's Mutational Signature Framework \citep{alexandrov2020repertoire, gori2018sigfit, diazgay2023SigProfilerAssignment}. \texttt{SigProfiler} was the first published framework for de novo inference of mutational signatures and its application led to the COSMIC mutational signatures catalogue \citep{alexandrov2020repertoire}.

Building upon the foundations of \texttt{SigProfiler}, \texttt{SigFit} is an R package that offers a Bayesian approach to mutational signature analysis \citep{gori2018sigfit}. It allows for a probabilistic framework with various prior distributions (e.g., Gamma and Dirichlet) on the signatures ($P$) and exposures ($E$) matrices. The model has options of Multinomial, Poisson, Negative Binomial, or Normal for the likelihood on the mutation counts matrix ($M$) \citep{gori2018sigfit}. Other computational packages include \texttt{SignatureAnalyzer} \citep{kasar2015whole, kim2016somatic}, \texttt{SigneR} \citep{rosales2016signeR}, \texttt{SigFlow} \citep{wangSigFlow2020}, \texttt{compNMF} \citep{zito2024compressivebayesiannonnegativematrix}, and \texttt{bayesNMF} \citep{landy2025bayesnmf}. These all perform de novo extraction of signatures, but the implementations differ.

The open-source R package \texttt{bayesNMF} implements the Normal-Truncated Normal model for mutational signature analysis and includes built-in convergence criteria and visualization capabilities for plotting estimated signatures. In our work, we used \texttt{bayesNMF} as the foundation upon which we extended the model to incorporate a Truncated Multivariate Normal prior for $P$ and developed a hierarchical framework that learns the covariance structure among mutation types. We leveraged the Gibbs updates for parameters other than the $P$ matrix, the convergence criteria, and plotting capabilities of the original package as a starting point for our model and code.

\subsection{Dependence Structure in Signatures Matrix}

Existing Bayesian NMF models for mutational signatures assume independence across mutation type probabilities across signatures \citep{rosales2016signeR}. That is, the prior assumes the rows of the $P$ matrix are independent. There is reason to believe, however, that there should exist some dependence structure across mutation type probabilities.

Some SBS mutations are more similar than others. For example, in the alphabet of 96 mutation types, $A[C>A]A$ may be more similar to $A[C>T]A$ than it is to $T[T>G]T$. Mutations with similar sequence contexts often co-occur in the same mutational processes, and these dependencies can reflect underlying biological mechanisms. For example, the APOBEC family of cytidine deaminases, which are known to play a role in various cancers, preferentially deaminate cytosines in specific nucleotide contexts such as TpCpN trinucleotides, where N represents any nucleotide \citep{dananberg2024apobec}. This activity results in characteristic C>T and C>G mutations primarily in these targeted contexts, giving rise to distinct mutational signatures such as SBS2 and SBS13. These signatures frequently co-occur in regions of localized hypermutation, known as kataegis, further emphasizing the role of sequence context in driving mutational patterns. 

To account for these dependencies, a few models have been developed that explicitly consider the sequence context of mutations. \citet{shiraishi2015simple} introduced a framework that treats the left context, right context, and center SNP as separate features. This way, mutations with the same center base substitution or the same context are explicitly similar in terms of their features. Expanding on this idea, \citet{zhang2020cancer} utilized a machine learning-based approach inspired by natural language processing (NLP) to represent longer contexts surrounding mutations. By decomposing features from extended contexts and learning embeddings, their method captures subtle dependencies and patterns that are otherwise overlooked by traditional approaches.

Considering these dependencies provides a more nuanced representation of the $P$ matrix, offering deeper insights into the biological processes underlying cancer mutations. To address the limitation of current Bayesian NMF methodologies, this thesis builds on existing Bayesian NMF methodologies to introduce a novel approach that explicitly incorporates the dependence structure across mutation type probabilities into the modeling framework. Additionally, this project introduces a novel hierarchical modeling framework, allowing the covariance structure of the $P$ matrix to be learned adaptively rather than specified a priori. We also contribute to the existing open-source R software package \texttt{bayesNMF} by incorporating our methodology, available at \href{https://github.com/88il/corBayesNMF}{88il/corBayesNMF}. To exemplify the accuracy and efficiency of our methodology, we conduct simulation studies as well as apply it to a cancer type from the Pan Cancer Analysis of Whole Genomes (PCAWG) database \citep{ICGC-TCGA-2020-PCAWG}.

\section{Data}

\subsection{COSMIC Data}

The Catalogue of Somatic Mutations in Cancer (COSMIC) provides a reference compendium of mutational signatures derived from large-scale cancer sequencing studies. Pioneering work by Alexandrov and colleagues applied non-negative matrix factorization (NMF) to somatic mutation data from over 30 cancer types, uncovering distinct recurrent mutation patterns now recorded as COSMIC mutational signatures \citep{alexandrov2013deciphering}. These single-base substitution (SBS) signatures represent specific DNA mutation processes operative in tumors, each identified by characteristic patterns of base changes across the cancer genome \citep{hu2020characteristics}.

As discussed in Section \ref{sec:Mutational Signatures}, somatic mutations are categorized into 96 SBS mutation types. Each COSMIC SBS signature is essentially a probability distribution across these 96 mutation types. In other words, each signature is defined by the proportion of each of the 96 mutation types it comprises. 

COSMIC SBS signatures are typically labeled SBS1, SBS2, SBS3, etc., and many have known or suspected etiologies. For instance, SBS1 is a “clock-like” signature linked to age-related 5-methylcytosine deamination, SBS4 is caused by tobacco smoking, and SBS7a/b are caused by UV light exposure (pyrimidine dimer formation) \citep{alexandrov2016mutational}. Each signature’s pattern is thought to reflect a particular mutagenic process or DNA repair failure. These reference signatures, though not ground truth, are widely used to interpret cancer genomes. By examining which COSMIC signatures appear in a tumor, one can infer the mutational processes that have acted on that tumor \citep{park2023diffsig}.

The COSMIC data (v3.3.1) \citep{tate2019cosmic} that we use is imported directly from the Sanger Institute's repository and formatted into a matrix where each column represents a normalized SBS signature \citep{alexandrov2020repertoire}. We use the COSMIC signatures in this thesis both as a reference and as a basis for simulation studies.

\subsection{PCAWG}

The Pan-Cancer Analysis of Whole Genomes (PCAWG) project is an international consortium effort that generated an expansive dataset of somatic mutations by sequencing whole genomes of primary cancers. In total, PCAWG analyzed 2,658 whole-cancer genomes across 38 tumor types, providing one of the most comprehensive surveys of genomic alterations in cancer to date \citep{ICGC-TCGA-2020-PCAWG}.

For this thesis, we utilize mutation count matrices derived from donor-centric, US-only PCAWG data, as made available through the UCSC Xena browser.  The original dataset encompassed 23 different histology groups corresponding to distinct cancer types. To ensure robustness, groups with fewer than 10 samples were excluded, resulting in 20 datasets being retained for further analysis. For each histology group, a separate mutational count matrix was built based on the 96 SBS types as built by Landy et al. (2025) \citep{landy2025bayesnmf}.

Hypermutated samples, tumors with an exceptionally high number of mutations, often due to defects in DNA repair mechanisms such as mismatch repair deficiency—are excluded from our analysis \citep{landy2025bayesnmf}. These hypermutated samples can drive de novo signature extraction such that hypermutation-specific signatures are incorrectly attributed to the broader tumor population \citep{maura2019practical}. By excluding hypermutated samples, we ensure that the estimated signatures accurately reflect the mutational patterns characteristic of the majority of tumors.

This thesis uses these curated count matrices for the application of our novel methods which are incorporated into the existing \texttt{bayesNMF} package. In our implementation, we extend the original methodology with a TruncMVN prior and a Bayesian hierarchical model. For the application of the methods to the PCAWG data, the rank (the number of distinct mutational signatures) is fixed based on the optimal value as determined by Landy et al.'s Sparse Bayesian Factor Inclusion (SBFI) method \citep{landy2025bayesnmf}.

\section{Methods}

\subsection{Truncated Multivariate Normal (MVN) Model}

While there are existing implementations of the Normal-Truncated Normal model, such as in \texttt{bayesNMF} or \texttt{SignatureAnalyzer}, to our knowledge there has been no implementation of the method where the signatures are modeled as a Truncated Multivariate Normal, as opposed to many Truncated Univariate Normals. In this section, we will present our model that takes into account the dependence structure across mutation type probabilities.

\subsubsection{Theoretical Framework}

The MVN Truncated Normal and Truncated Normal are used as prior distributions on $P$ and $E$, respectively, and their non-negative truncation region are chosen to ensure that $P$ and $E$ matrices are non-negative. Additionally, just as the Normal-Truncated Normal model has nice conjugacy, our Normal-MVN Truncated Normal model also has nice conjugacy. That is, the full conditionals of columns of the P matrix remain MVN Truncated Normal. Our model incorporates a covariance matrix $\Sigma$ into the parameterization of the $P$ matrix, resulting in it becoming a Multivariate Truncated Normal. The $E$ matrix remains a Univariate Truncated Normal. Finally, the likelihood ($M$ matrix) remains Univariate Normal, but using matrix notation, we rewrite the $M$ matrix to take on the form of a Multivariate Normal distribution with a diagonal covariance matrix.

We now formalize the model. We have $K$ mutation types, $G$ tumor samples, and $N$ signatures. The mutation counts matrix: $M \in \mathbb{R}_{\geq 0}^{K \times G}$, the signatures matrix: $P \in \mathbb{R}_{\geq 0}^{K \times N}$, the exposures matrix: $E \in \mathbb{R}_{\geq 0}^{N \times G}$.

Define $\Sigma^M \in \mathbb{R}^{K \times K}$ (the covariance matrix, across mutation types, of $M^T$) for $j, k \in \{1, \dots, 96\}$, to be the diagonal matrix:
\begin{align*}
    \Sigma_{jk}^M &= \begin{cases}
    \sigma_k^2 & \forall j = k  \\
    0 & \forall j \neq k 
\end{cases} \\
 &= \begin{bmatrix}
   \sigma_1^2 & 0 & \dots  \\ 
   0 & \ddots & \\ 
   \vdots &  & \sigma_{96}^2
 \end{bmatrix}
\end{align*}

So, we have $\vv{p_n} \in \mathbb{R}^{K \times 1}$ (column of $P$, n-th signature), $\vv{M_g} \in \mathbb{R}^{K \times 1}$ (column of $M$, g-th tumor sample). The likelihood is:
$$\vv{M_g} \sim \text{MVN}((\vv{PE})_g, \Sigma^M)$$

At the signatures, exposures, and data level, the priors are:
\begin{align*}
    \vv{p_n} &\sim \text{MVNTruncNorm}(\vv{\mu_n}^P, \Sigma, 0, \infty) \\
    E_{ng} &\sim \text{TruncNorm}(\mu_{ng}^E, \sigma_{ng}^{2E}, 0, \infty) \\
    \sigma_g^2 &\sim \text{Inverse-Gamma}(\alpha_g, \beta_g)
\end{align*}

The hyperpriors are:
\begin{align*}
    \mu_{kn}^P &\sim \text{Normal}(m_{kn}^P, s_{kn}^P) \\
    \mu_{ng}^E &\sim \text{Normal}(m_{ng}^P, s_{ng}^P) \\
    \sigma_{ng}^{2E} &\sim \text{Inverse-Gamma}(a_{ng}^E, b_{ng}^E)
\end{align*}

The full conditional distributions for $E$ and $\sigma_k^2$ remain unchanged from the Normal-Truncated Normal model (see Appendix~\ref{Appendix: Norm-Trunc Norm Posterior Derivation}). The prior for $\vv{p_n}$ is conjugate to the Normal likelihood, so its full conditional remains distributed as a Truncated MVN distribution. In particular, the full conditional is distributed as:

$$p(\vv{p_n} | P_{(-n)}, E, M, \sigma^2) \sim \text{MVNTruncNorm}(\vv{\mu_n}^{new}, \Sigma^{new}, 0, \infty)$$ 

where
\begin{align*}
    \vv{\mu_n}^{new} &= A^{-1}B = \left(\Sigma^{-1} + (\sum_g E_{ng}^2)(\Sigma^M)^{-1}\right)^{-1} \left(\Sigma^{-1}\vv{\mu_n}^P + (\Sigma^M)^{-1} \vv{A^*}\right) \\
    \Sigma^{new} &= A^{-1} = \left(\Sigma^{-1} + (\sum_g E_{ng}^2)(\Sigma^M)^{-1}\right)^{-1}
\end{align*}

and 
$$A_k^* = \sum_g E_{ng} (M_{kg}-\hat{M}_{kg(-n)})$$
$$\hat{M}_{kg(-n)} = \sum_i P_{ki}E_{ig} - P_{kn}E_{ng}$$

(For the detailed full conditional derivation, see Appendix~\ref{Appendix: MVN-Truncated MVN derivation}.) 

Intuitively, we see that the full conditional mean of $P$ is a weighted combination of the prior mean on $P$ ($\vv{\mu}_n^P$) as well as the residuals of the decomposition excluding signature $n$ ($M_{kg} - \hat M_{kg(-n)}$). The $A_k^*$ term in the mean vector can be interpreted as a linear combination of weights $E_{ng}$ and residuals after accounting for other signatures $(M_{kg}-\hat{M}_{kg(-n)})$. 

So, in the full conditional, the mean can be interpreted as the sum of the prior mean inverse weighted by the prior variance and the mean based on the data–what is left unexplained (residuals) inverse weighted by the variance of the data.

\subsubsection{Comparison of Posterior to nonMVN Model}

Substituting in the identity matrix for $\Sigma$, the covariance matrix of $P$, in the above section aligns with the full conditional distribution for the Norm-TruncNorm model (which we will call the ``nonMVN" model), as expected.

Furthermore, the updates for $E_{ng}$ and $\sigma_k^2$ remain unchanged in the MVN model as compared to the nonMVN model.

Specifically, using the same notation as above, the prior specifications of the nonMVN method are:
\begin{align*}
    P_{kn} &\sim \text{TruncNorm}(\mu_{n}^P, \sigma_{kn}^{2P}, 0, \infty) \\
    E_{ng} &\sim \text{TruncNorm}(\mu_{ng}^E, \sigma_{ng}^{2E}, 0, \infty) \\
    \sigma_g^2 &\sim \text{Inverse-Gamma}(\alpha_g, \beta_g)\\
    M_{kg} &\sim \text{N}((PE)_{kg}, \sigma_k^2)
\end{align*}

The full conditional distributions leave $P_{kn}$ distributed as TruncNorm, $E_{ng}$ distributed as TruncNorm, and $\sigma_k^2$ distributed as InvGamma. For details, see Appendix~\ref{Appendix: Norm-Trunc Norm Posterior Derivation}.

\subsubsection{MCMC Algorithm}

Our goal is to maximize the joint posterior:
$$p(P, E, \sigma^2 | M)$$

Sampling directly from this high-dimensional posterior distribution is not tractable, and maximizing the posterior distribution with coordinate gradient descent, while possible, does not allow for uncertainty quantification. Instead, we leverage the fact that each of the full conditional distributions is tractable. This allows us to use a Gibbs sampler—a type of Markov chain Monte Carlo (MCMC) algorithm—to obtain samples from the joint posterior.

Gibbs sampling works by iteratively updating each parameter (or block of parameters) conditional on the current values of all other parameters. In each iteration, we sample from the full conditional distributions:
\begin{enumerate}
    \item for each n: update $\vv{p_n} \sim p(\vv{p_n} | M, P_{(-n)}, E, \sigma^2, \Sigma)$
    \item for each n, g: update $E_{ng} \sim p(E_{ng} | M, P, E_{(-ng)}, \sigma^2)$
    \item update $\sigma^2 \sim p(\sigma^2 | M, P, E)$
\end{enumerate}

until convergence. Full details on the convergence control used and the implementation of the sampling from the full conditional are available in the Supplementary Materials.

\subsubsection{Choice of Prior Covariance of Signatures Matrix}

In our Bayesian framework, we modify the \texttt{bayesNMF} function to accept an externally supplied correlation matrix of the signature matrix $P$. Specifically, we use the James–Stein Shrinkage estimate of the COSMIC (v3.3.1) correlation matrix as the input for our simulation studies \citep{Heumann2011JS, tate2019cosmic}. This prior leverages well-established mutational signatures while not being overly specific to any one cancer type. The data-driven shrinkage mechanism regularizes the sample correlation matrix by pulling it toward a structured target (the identity matrix), thereby reducing estimation variance and ensuring that the resulting matrix is positive definite. 

Mathematically, the James–Stein shrinkage estimator for a covariance matrix can be expressed as a convex combination of the sample estimate $S$ and a target matrix $T$:
$$\hat{\Sigma} = (1-\lambda)S + \lambda T$$

where $\lambda \in [0,1]$. Ledoit and Wolf (2003) suggest an analytic formula for estimating $\lambda$ such that the Mean Squared Error (MSE) of $\hat{\Sigma}$ is minimized \citep{ledoit2003improved}.

This is particularly important in our setting where the number of variables (mutation types) exceeds the number of observations (tumor samples), leading to a singular sample correlation matrix. Such situations of ``small n, large p" are often encountered in bioinformatics and statistical genomics \citep{schaffer2005shrinkage}. By using the \texttt{cov.shrink()} function from the \texttt{corpcor} package, followed by a conversion to a correlation matrix via \texttt{cov2cor()}, we obtain a robust, well-conditioned prior that can be safely inverted and used in subsequent Bayesian inference steps \citep{R-tmvtnorm2023}.

In our implementation, $\lambda \approx 0.0991$ which indicates that the sample correlations are only slightly adjusted—about $9.91\%$ of the way toward the target. So, most of the structure of the original sample correlation matrix is retained, but it is regularized slightly to ensure that the matrix is positive definite.

\subsection{Hierarchical Model}

In this section, we discuss our novel Bayesian Hierarchical methodology for mutational signature analysis that aims to learn the covariance structure of the signatures matrix, rather than being specified upfront. The model aims to learn the dependence structure of the signatures to allow for a better understanding of biological interactions and how these change across different cancer types.

\subsubsection{Model Specification and Priors}
\label{sec:hierarchical model specification and priors}

For the 3-parameter hyperprior model, we assume the same Normal-Truncated MVN model as detailed in our model from the previous section. However, now we have an additional level of hyperpriors imposed on the covariance matrix of $P$.

We parameterize the $96 \times 96$ covariance matrix of $P$ to reflect two types of correlation: correlation between mutation types with the same center mutation ($\rho_{\text{same}}$) and correlation between mutation types that do not have the same center mutation ($\rho_{\text{diff}}$). That is:
$$\Sigma_{i,j} = \begin{cases}
    \sigma_P^2 &{\text{if i=j}} \\
    \rho_{\text{same}}\cdot \sigma_P^2 &\text{if mutation type i and mutation type j share the same center mutation} \\
    \rho_{\text{diff}} \cdot \sigma_P^2 &{\text{else}}
\end{cases}$$

This structure assumes a constant base variance across mutations. This simplistic assumption gives a more parsimonious representation and ensures that we can focus on learning the correlation structure.

For the prior distributions on these parameters, we set:
\begin{itemize}
    \item Base Variance: $\sigma_P^2 \sim \text{Inverse-Gamma}(\alpha, \beta)$
    \item Hyperpriors: $\rho_{\text{same}} \sim \text{Beta}(\alpha_{\text{same}}, \beta_{\text{same}}), \;\; \rho_{\text{diff}} \sim \text{Beta}(\alpha_{\text{diff}}, \beta_{\text{diff}})$
\end{itemize}

In our implementation, we set all six parameters $\alpha, \alpha_{\text{same}}, \alpha_{\text{diff}}, \beta, \beta_{\text{same}}, \beta_{\text{diff}}$ to be equal to $2$. When calling the \texttt{bayesNMF} function, the user can specify these six parameters themselves if they want to adjust the prior. Here, we set fairly vague and uninformative priors so that $\rho_{\text{same}}, \rho_{\text{diff}} \in [0,1]$, centered around 0.5. We then rescale $\rho_{\text{diff}}$ after sampling from the Beta distribution so that it lies in the range $[-1,1]$. This choice was informed by our exploratory analysis which suggested that $\rho_{same}$ is likely nonnegative, but $\rho_{diff}$ is not necessarily nonnegative. As such, mapping $\rho_{diff}$ from $x \to 2x-1$ would result in the transformation having a distribution with desired support $[-1,1]$ and a mode at zero.

However, different mutation types may have different variances so it can be worthwhile to account for heteroskedasticity with an alternative parameterization. Alternatively, we considered a Half-Normal or Half-Cauchy for the Base Variance as well as an LKJ prior or Inverse-Wishart prior for the full correlation matrix. The LKJ prior is a more complex, full correlation matrix parameterization. The Inverse-Wishart prior is a generalization of the Inverse-Gamma distribution and fully parameterizes a covariance matrix. It is commonly used for covariance matrices and is a conjugate prior for the covariance matrix of a multivariate normal distribution, but it can be overly restrictive and lacks interpretability for correlation structures. Specifically, we would have: $\Sigma \sim \text{Inverse-Wishart}(\psi, \nu)$ where $\psi$ is a scale matrix and $\nu$ is the degrees of freedom. In our method, we prioritize simplicity and interpretability, but such alternatives can be explored in future works.

\subsubsection{Implementation: Metropolis-Hastings Random Walk}

In our Hierarchical model, because we are simply adding an additional layer of parameters ($\sigma_P^2$, $\rho_{\text{same}}$, $\rho_{\text{diff}}$) on top of our existing MVN model, we will still perform the same the same Gibbs updates for $P, E, \sigma^2$ sampling from their full conditional distributions. However, before sampling $P$, we now must update the three parameters $\sigma_P^2, \rho_{\text{same}}, \rho_{\text{diff}}$ by sampling from their full conditional distributions and construct a new $\Sigma$ using the resampled parameters. This new $\Sigma$ will be used in the sampling of $P$. Each iteration consists of the following steps:
\begin{enumerate}
    \item update $\sigma_P^2, \rho_{\text{same}}, \rho_{\text{diff}}$
    \item for each n: update $P_{-\vv{n}} \sim p(\vv{p_n} | M, P_{-\vv{n}}, E, \sigma^2, \sigma_P^2, \rho_{\text{same}}, \rho_{\text{diff}})$
    \item for each n, g: update $E_{ng} \sim p(E_{ng} | M, P, E_{-ng}, \sigma^2)$
    \item update $\sigma^2 \sim p(\sigma^2 | M, P, E)$
\end{enumerate}

The three parameters that parameterize the covariance matrix of $P$, do not have closed-form full conditional distributions. So, to update $\sigma_P^2, \rho_{\text{same}}, \rho_{\text{diff}}$, we employ a Metropolis-Hastings (MH) random walk algorithm. In the algorithm, a proposal for the parameter value that is being updated is generated by drawing from a normal distribution centered at the current value of the parameter with a specified standard deviation (\texttt{proposal\_sd}). The proposal is then checked against predefined lower and upper bounds ($[1\mathrm{e}{-6}, 100]$ for $\sigma_P^2$, $[0, 1]$ for $\rho_{\text{same}}$, $[-1, 1]$ for $\rho_{\text{diff}}$). If the proposed value lies outside the interval, it is immediately rejected, and the current value is retained. Otherwise, we compute the unnormalized log posterior densities at both the current and proposed values. The acceptance ratio is calculated as the exponential of the difference between the proposed and current log posterior values. Finally, a uniform random draw determines whether to accept the proposed value, thus updating the parameter.

The proposal standard deviation is set to $0.1$ for all parameters by default, but should be tuned in some cases. We want to ensure a moderate step size that maintains a reasonable acceptance rate. This MH random walk provides the flexibility needed for parameters without analytic Gibbs updates while integrating into our overall MCMC scheme.

\subsubsection{Strengths and Weaknesses of the Model}

Our hierarchical model offers a parsimonious representation of the mutational signatures by parameterizing the covariance structure with only a few key parameters ($\sigma_P^2, \rho_{\text{same}}, \rho_{\text{diff}}$). This approach enhances interpretability and also reduces the risk of overfitting, especially in scenarios with limited data. Moreover, the \texttt{build\_covariance\_matrix} function and overall implementation of our method is designed to be flexible and easily extendable. In particular, the framework can be extended to larger alphabets by incorporating additional surrounding contexts, which is particularly useful as researchers seek to analyze more detailed mutational patterns. The model’s integration into the \texttt{bayesNMF} framework also allows it to leverage existing robust Bayesian inference methods while introducing the ability to learn the covariance structure directly from the data.

Despite its advantages, our model does have limitations. The assumption of a single $\rho_{\text{same}}$ parameter for all mutation types may be an oversimplification as it might not capture all the nuances in the correlation structure observed in real data. Additionally, the algorithm can be computationally intensive. The MCMC sampling procedure, particularly with the truncated MVN and Metropolis-Hastings random walk updates, often requires a long runtime to achieve convergence in high-dimensional settings. Lastly, tuning the proposal standard deviation remains challenging. While we have set it to $0.1$ by default to ensure a moderate step size and reasonable acceptance rate, the optimal value may vary across parameters and datasets, necessitating further experimentation and fine-tuning for robust performance.

\section{Results}

\subsection{Dependence Structure in Signatures Matrix}

Currently, NMF methods for mutational signature analysis assume independence across mutation type probabilities in the signatures matrix. We present both biological and statistical analysis of the dependence structure in the signatures matrix ($P$) so as to motivate the necessity to model the covariance of the $P$ matrix. 

\subsubsection{Biological Motivation}

We begin by examining the $96 \times 96$ correlation matrix of the COSMIC $96 \times 79$ signatures matrix. In doing so, we can get a sense for what the correlations between pairs of mutation types look like.

\begin{figure}
\includegraphics[width=\textwidth]{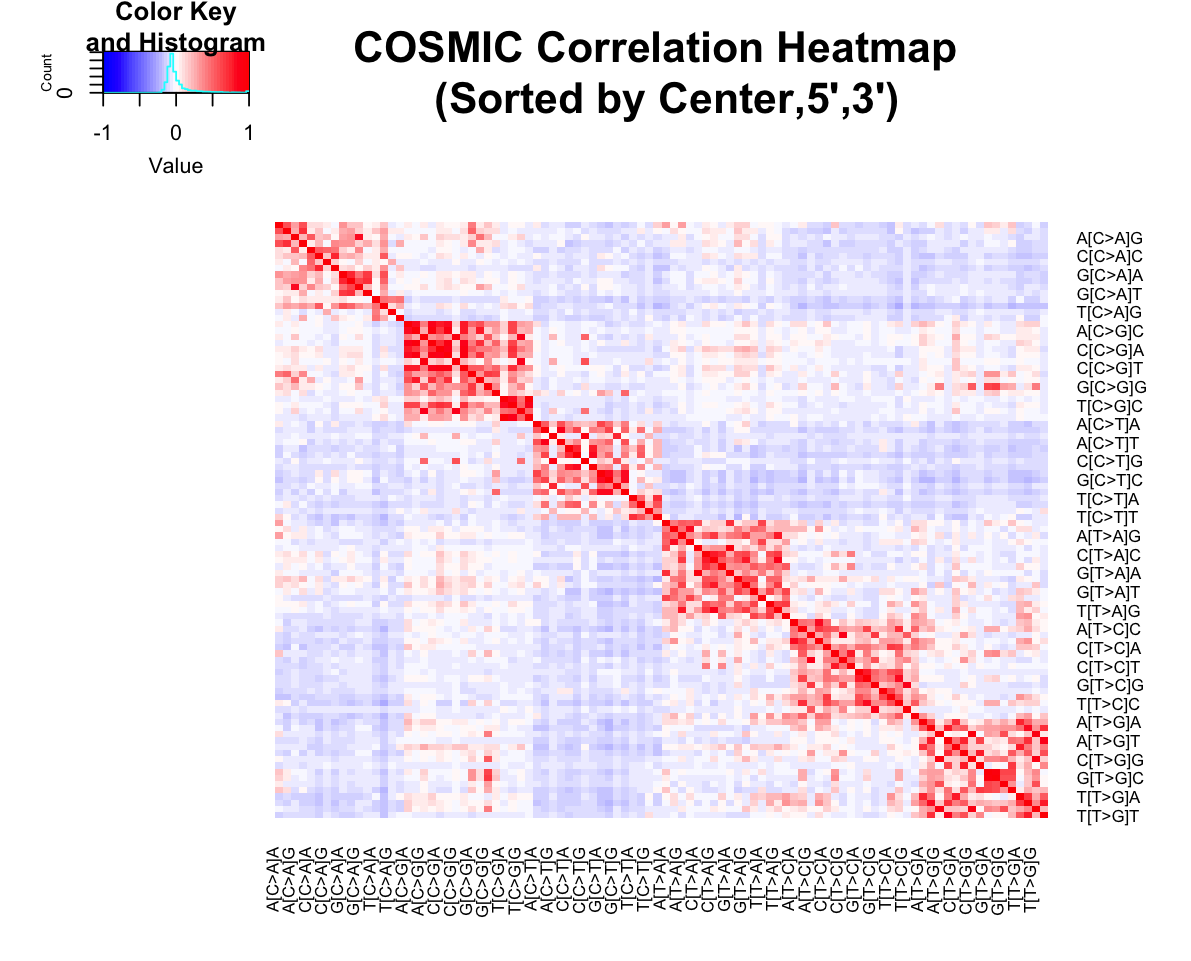}
\caption{Heatmap of COSMIC correlations between pairs of the 96 mutation types, sorted by center, left, and right mutation.}
\label{fig:cosmicheatmap}
\end{figure}

The heatmap in Figure~\ref{fig:cosmicheatmap} has mutation types on the rows and columns sorted first by center mutation type (C>A, C>G, C>T, T>A, T>C, T>G), then by the left base (A, C, G, T), and finally by the right base (A, C, G, T). From the heatmap, we can see six distinct clusters along the diagonal. These six clusters correspond to the six center mutation types, and as expected, we observe strong, positive correlation in these clusters. Additionally, there appears to be some structure outside of the six main clusters on the diagonal; however, the correlations are generally weaker and more negative. 

\begin{figure}[htbp]
  \centering
  \begin{subfigure}[b]{0.45\textwidth}
      \centering
      \includegraphics[width=\textwidth]{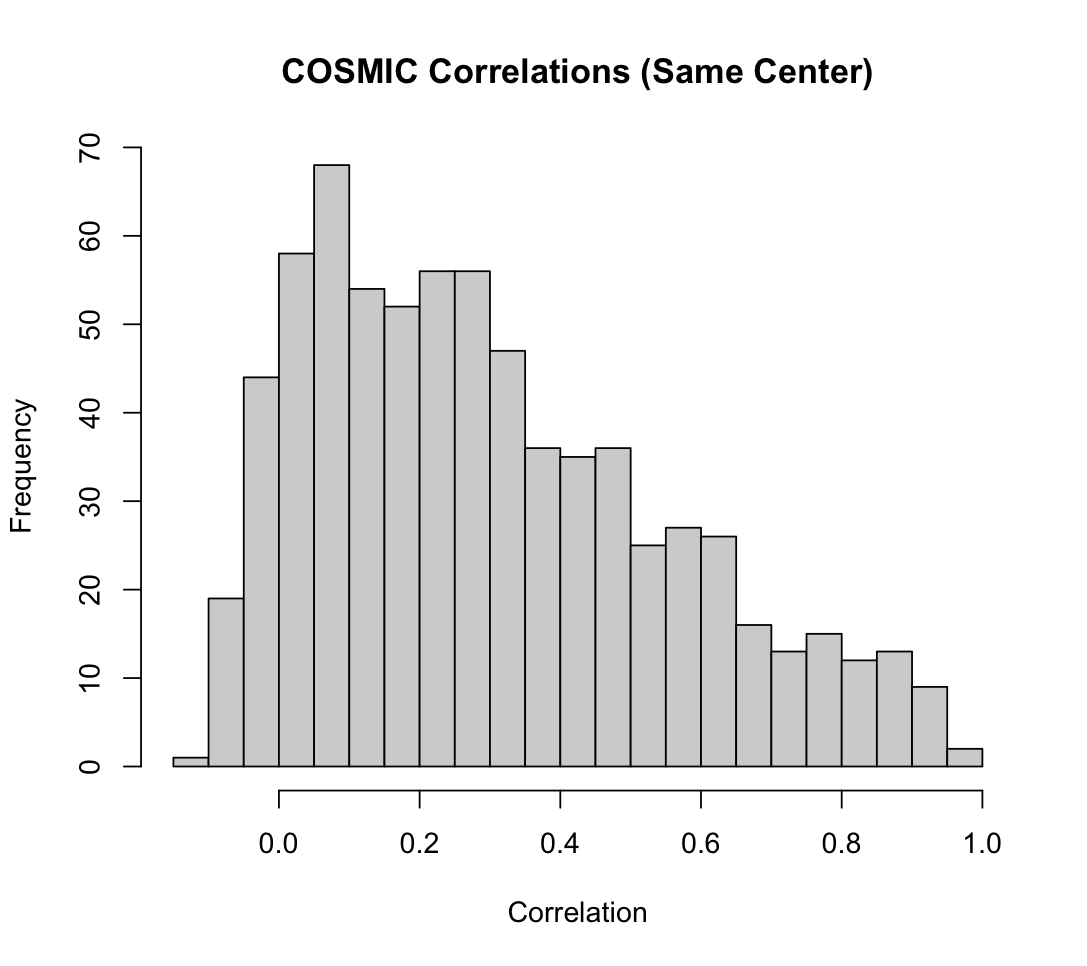}
      \label{fig:cor_samecenter_hist}
  \end{subfigure}
  \hfill
  \begin{subfigure}[b]{0.45\textwidth}
      \centering
      \includegraphics[width=\textwidth]{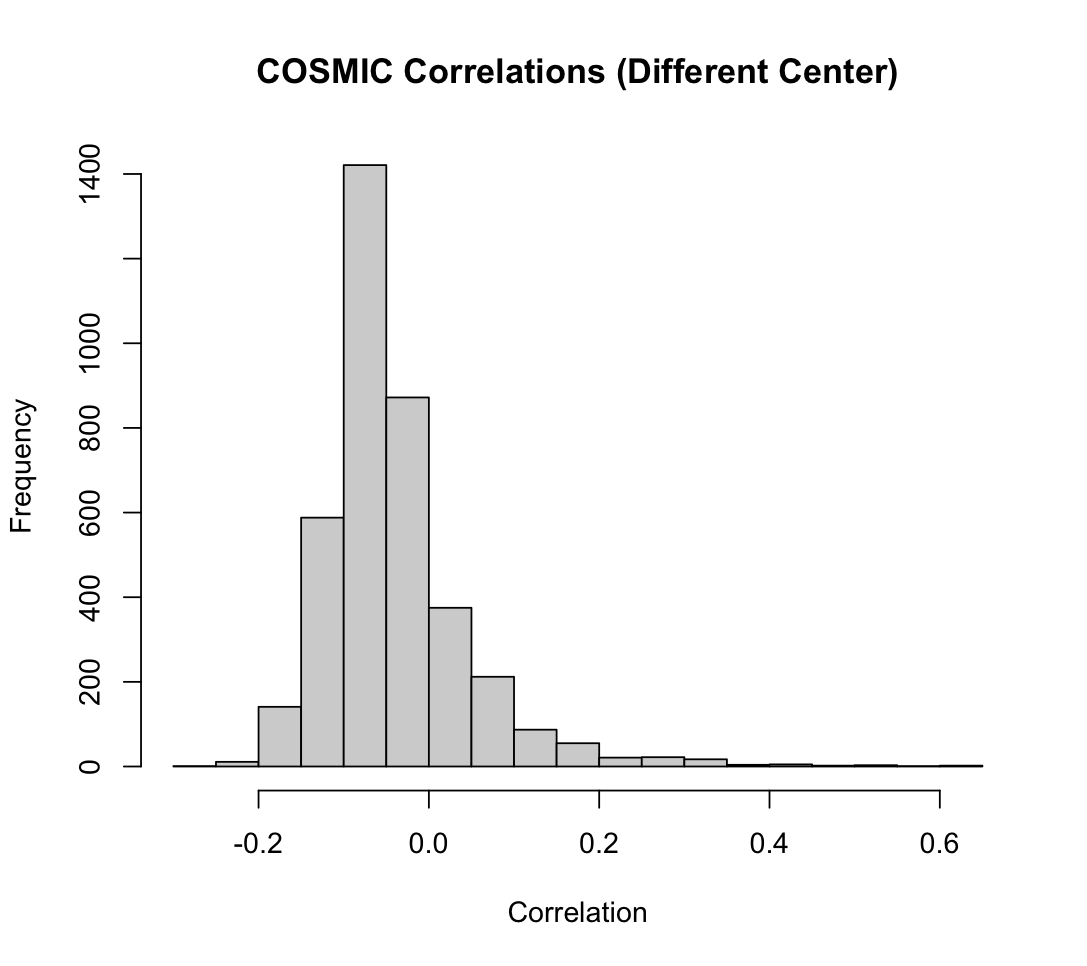}
      \label{fig:cor_diffcenter_hist}
  \end{subfigure}
  \caption{Distribution of COSMIC correlations for pairs of mutation types with the same and with different center mutations.}
  \label{fig:cosmic_cor_histograms}
\end{figure}

Figure~\ref{fig:cosmic_cor_histograms} shows the distribution of the correlations for the elements in the six clusters (excluding the 96 diagonal entries which are all equal to one and excluding the upper triangle so each unordered pair appears exactly once), giving a total of $\frac{1}{2} \cdot (16\times16\times6-96)=720$ correlations, as well as the distribution of the correlations for the 3840 pairs of mutation types that do not have the same center mutation. The same center histogram contains values ranging from $-0.11$ to $0.98$ with a median correlation of $0.26$ and a mean correlation of $0.30$ while the different center histogram contains values ranging from $-0.27$ to $0.65$ with a median correlation of $-0.06$ and a mean correlation of $-0.04$. 

As such, through this preliminary examination of the correlation structure in the COSMIC signatures matrix, we indeed see evidence of nonzero correlation. That is, it does not seem that mutation types are independent from each other. Further, this breakdown of correlations between pairs of mutation types that share the same center mutation versus those that do not share the same center mutation will be informative for our hierarchical model.

\subsubsection{Number of Significant Correlations is Higher Than Expected By Chance Alone}

To more robustly exemplify this dependence between the mutation types, we now show that the number of significant correlations between mutation types is higher than expected by chance alone. In particular, we perform a permutation test to evaluate whether the correlations between COSMIC signatures are statistically significant, assuming independence between mutation types under the null hypothesis.

Under the null hypothesis, signatures have independent mutation types, so the first step in generating a null distribution is to model each of the 96 mutation types as a univariate truncated normal random variable (independent from each other). Formally, let $P^0_{kn}$ represent the proportion of mutations for mutation type $k$ in signature $n$. For a given mutation type $k$ under the null hypothesis, we assume:
$$P^0_{kn} \sim \text{TruncNorm}(\mu_k, \sigma_k, 0, \infty) \; \; \forall n = 1, \dots, 79$$

Here, the scalars $\mu_k$ and $\sigma_k$ are the mean and standard deviation, respectively, of the Truncated Normal distribution, and the lower and upper bounds (truncation region) are $0$ and $\infty$ due to the fact that each element $P^0_{kn}$ of the signatures matrix must be nonnegative. Upon specifying this model, we use the \texttt{fitdist} function on the COSMIC signatures matrix to obtain the Maximum Likelihood Estimates (MLE) for ($\mu_k, \sigma_k$) for mutation type $k$ \citep{fitdist}.

Now that we have our null distribution assuming independent mutation types in P, we simulate \texttt{n\_repetitions}=1000 signature matrices ($96 \times 79$) from the null distribution. Specifically, for each simulated signature matrix, we need to sample a $96 \times 1$ vector of probabilities (for each mutation type) from the null truncated normal distribution we derived. We then normalize these signatures to sum to one in order to represent proportions. We sample \texttt{n\_signatures}=79 random mutational signatures in this fashion to construct one simulated signature matrix.

For each of the signature matrices, we compute the simulated correlation matrix ($96 \times 96$) across mutation type probabilities. This gives a null distribution for the correlation between any pair of mutations ($96 * 96 = 9216$ total pairs). Now, we compute p-values. Let $\Sigma$ be the correlation matrix of COSMIC, and $\Sigma^{0,(r)}$ be the correlation of the r-th simulated correlation matrix. Then we compare the observed correlation $\Sigma_{ij}$ to the simulated null distribution $\Sigma^{0,(r)}_{ij}$ for $r = 1,...,1000$. The p-value we obtain is $\frac{1}{R}\sum_{r = 1}^T I(|\Sigma^{0,(r)}_{ij}| > |\Sigma_{ij}|)$ for any pair of mutation types $(i,j)$.

We compute the proportion of sampled null correlations that are as extreme or more extreme (in absolute value) than the observed correlation to obtain the p-value for a given mutation type pair, and we do this for all $96\times96$ pairs. In other words, this p-value represents the probability that we observe something as extreme or more extreme than the COSMIC correlation by chance.

\begin{figure}
\includegraphics[width=\textwidth]{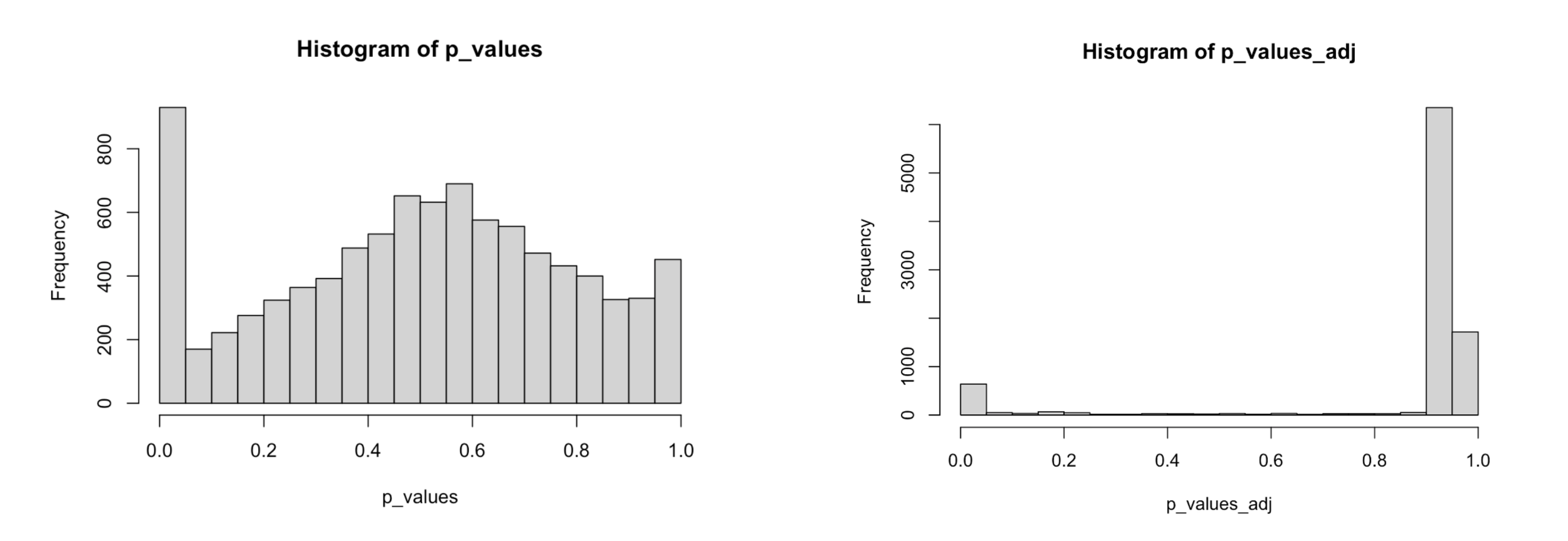}
\caption{p-values before and after Benjamini-Hochberg adjustment.}
\label{fig:pval_permutation_test}
\end{figure}

We use the Benjamini-Hochberg (BH) method to control for False Discovery rate (FDR) \citep{benjamini1995controlling}. This ranks the p-values in ascending order and adjusts each p-value based on its rank and the total number of tests, $G$. The adjusted p-value for the $i$-th p-value is:
$$\mathrm{adjusted\_pval}_i = \frac{\mathrm{pval}_i \cdot G}{\mathrm{rank}_i}$$

Based on these adjusted p-values (see Figure~\ref{fig:pval_permutation_test}), we can identify significant correlations at a significance level of $\alpha = 0.05$. With this BH correction, it is guaranteed that the FDR is at most $\alpha = 0.05$. Of the 9216 mutation type pair adjusted p-values, we obtain 638 adjusted p-values that are less than $\alpha = 0.05$. Consequently, the probability that a correlation is significant is:
$$\frac{638/2}{(9216-96)/2} = \frac{319}{4560} \approx 0.06996$$

which is larger than the expected threshold $\alpha = 0.05$, suggesting that the number of significant correlations exceeds what would be expected by chance alone under the null hypothesis of independence. Note that in the calculation, we must account for symmetry and diagonal elements. The numerator is halved since pair $(i,j)$ is counted twice: once as $(i,j)$ and once as $(j,i)$. In the denominator, we exclude the $96$ diagonal p-values when $i=j$ (since these p-values always equal $1$) before halving the quantity, too.

The higher-than-expected proportion of significant correlations implies potential biological relationships between mutation types, and these relationships could indicate shared mutational processes or pathways influencing specific mutation types.

\subsection{MVN Simulation Studies}

\subsubsection{Data}

For this simulation study, we generate a synthetic mutational counts matrix based on a subset of COSMIC SBS signatures. In our setting, we assume that there are $N=5$ true latent signatures and $G=60$ samples. The five signatures were randomly sampled from the COSMIC SBS v3.3.1 catalog to be SBS7d, SBS10b, SBS13, SBS17a, and SBS17b \citep{tate2019cosmic}. 

The $P$ (signatures) matrix is arranged such that each column corresponds to a signature and is normalized so that the elements of each column sum to 1.  This normalization allows each signature to be interpreted as a probability distribution over the 96 mutation types.

The exposures matrix $E$ is generated by sampling each of the $N \times G$ elements from an exponential distribution. That is, $E_{ij} \sim \text{Exponential}(0.001)$. The choice guarantees that exposures are strictly positive and can vary substantially across samples. Finally, we simulate each observed count $M_{kg}$ from a Poisson distribution:
$$M_{kg} \sim \text{Poisson}(\lambda_{kg} = (PE)_{kg})$$

\subsubsection{Results}

We compare the performance of our MVN method with the nonMVN method, in which independence across mutation types in the signatures matrix is assumed. Both methods specify a normal likelihood and truncated-normal prior. While we know the true rank is 5, when performing non-negative matrix factorization (NMF), the true rank is unknown. For the comparison of the MVN and nonMVN methods, we apply the methods on the dataset fixing the rank to be $N = 1, 2, \dots, 5$.

\begin{figure}
\includegraphics[width=\textwidth]{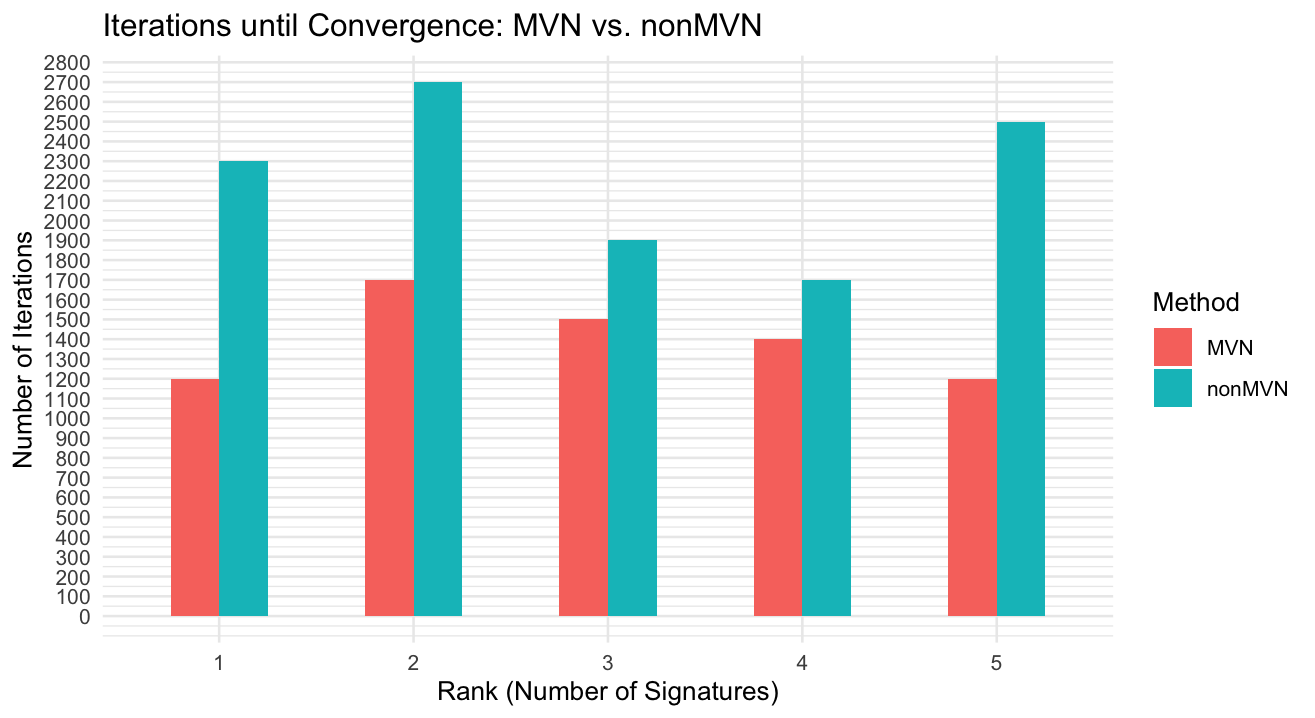}
\caption{Comparison of number of iterations until convergence using MVN and nonMVN methods for simulation study with sample size of G=60 and rank (number of signatures) of N=5.}
\label{fig:num_iter_mvn_nonmvn}
\end{figure}

\begin{figure}
\includegraphics[width=\textwidth]{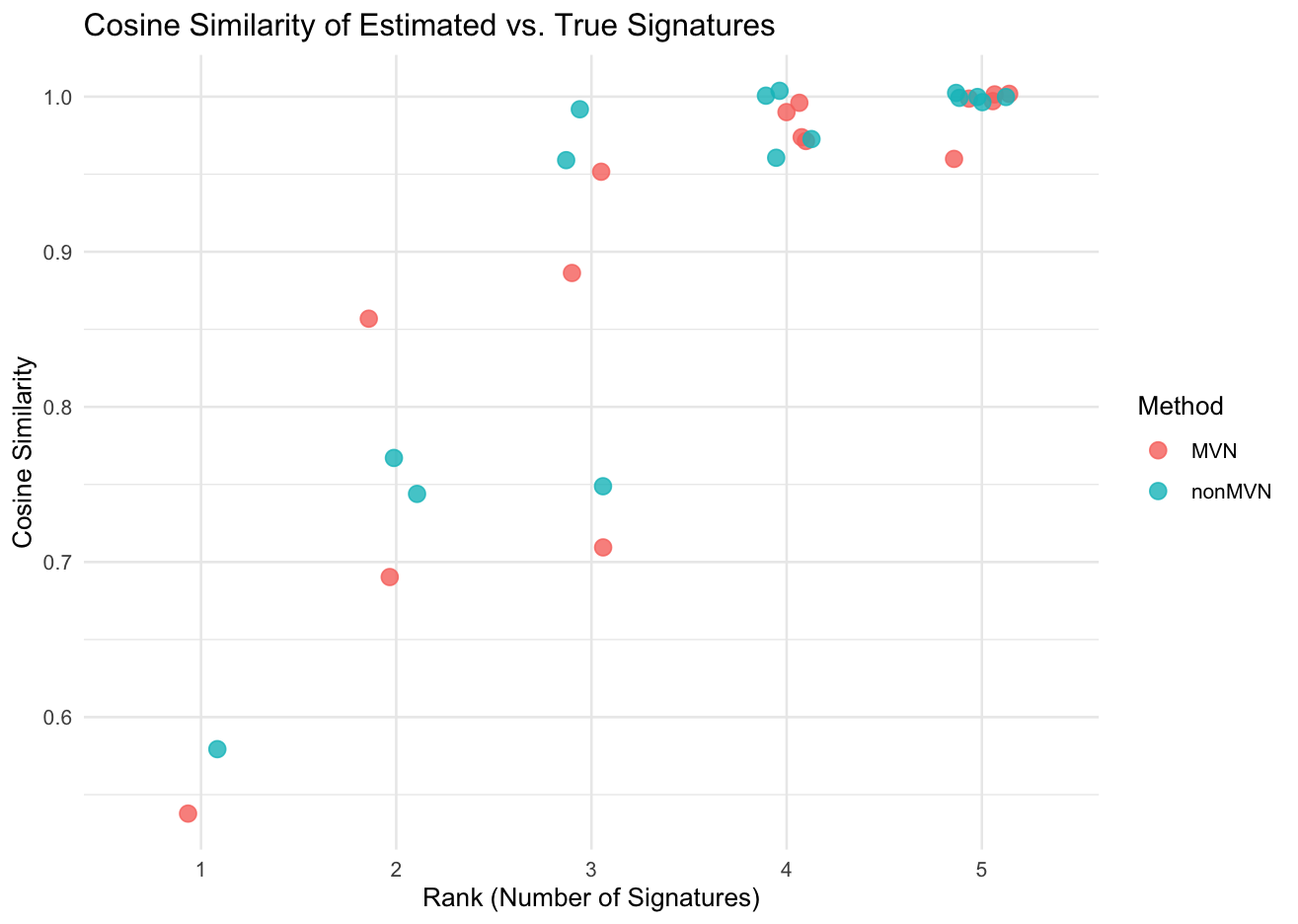}
\caption{Comparison of cosine similarities between estimated and COSMIC reference signatures using MVN and nonMVN methods for simulation study with sample size of G=60 and rank (number of signatures) of N=5.}
\label{fig:jitter_cossim}
\end{figure}

A notable strength of the MVN method in this simulation study is that it required fewer iterations until convergence than the nonMVN method. In Figure~\ref{fig:num_iter_mvn_nonmvn}, we see that the MVN model required at least 1000 fewer iterations in comparison to the nonMVN method for ranks of 1, 2, and 5, and required a couple of hundred fewer iterations for ranks of 3 and 4. Note that convergence is only checked for  every 100 samples. This means that a model showing 1200 iterations to converge implies convergence somewhere between 1101 and 1200.

Additionally, we see that both methods are comparable in terms of accuracy. To evaluate our estimates, we examine cosine similarity: a metric that quantifies the resemblance between two signatures (with a value of 1 indicating identical signatures) \citep{gori2018sigfit, alexandrov2013deciphering}. In Figure~\ref{fig:jitter_cossim}, we see that the cosine similarities of signatures between the two methods for any given rank that is specified are very close in value, which verifies the accuracy of our MVN method. The cosine similarity increases as the specified rank increases up until the true rank of five, at which point the cosine similarity for both models is $>= 0.96$ for all signatures.

While the number of iterations required until convergence is lower for the MVN model, the time needed for each iteration is larger. Specifically, the time for each iteration of the MVN model is on the scale of tenths of a second, while the time for each iteration of the nonMVN model is on the scale of hundredths of a second. This is expected as sampling a multi-variate truncated normal random variable is more complex and computationally intensive than sampling many univariate truncated normals.

Indeed, the per-iteration cost may be higher for the MVN model; however, the computational burden only really starts to pose an issue when scaling the method to more complex applications, for instance, if the alphabet is expanded or if the rank increases. Future optimizations that improve the sampling efficiency of the Truncated Multivariate Normal distribution could substantially reduce the overall runtime of the MVN method, thereby making the method more practical for large-scale analyses.

\subsection{Motivation for Hierarchical Model: MVN Improvement in Accuracy}

In the above simulation study, we saw that the MVN model took fewer iterations to converge compared to the nonMVN model. Both models were of similar accuracy for signature estimation, as seen through the cosine similarities. With a sample size of 60 for five latent signatures, it is reasonable that the nonMVN method already performs quite well in terms of accuracy. Similar simulation studies conducted by \citet{landy2025bayesnmf} show that generally, the larger the sample size is relative to the number of signatures, the more accurate the estimates will be as there is more data. 

For this simulation study, we generate several datasets with $N=8$ signatures and $G=10$ samples.  With a smaller sample size relative to the number of signatures, the nonMVN method is not able to perfectly estimate the signatures. Additionally, the signatures matrix used to generate these datasets was sampled using the 3-parameter specification of the covariance matrix defined in Section \ref{sec:hierarchical model specification and priors}, which poses a challenge for the nonMVN method since the nonMVN method assumes an identity matrix for the covariance matrix. 

Recall that in the 3-parameter hyperprior model, the covariance matrix of $P$, $\Sigma$, is parameterized as follows:
$$\Sigma_{i,j} = \begin{cases}
    \sigma_P^2 &{\text{if i=j}} \\
    \rho_{\text{same}}\cdot \sigma_P^2 &\text{if mutation type i and mutation type j share the same center mutation} \\
    \rho_{\text{diff}} \cdot \sigma_P^2 &{\text{else}}
\end{cases}$$

In this simulation study, we choose specific parameters: $\sigma_P^2=7, \rho_{\text{same}}=0.5, \rho_{\text{diff}} = -0.1$ in order to build our covariance matrix for $P$. Each column ($96 \times 1$) of $P$ is sampled from a Multivariate Truncated-Normal distribution using the Gibbs algorithm provided by the \texttt{tmvtnorm} package \citep{R-tmvtnorm2023}, with mean vector of the distribution set as the square root of the diagonal elements of $\Sigma$ to ensure that each component’s expected value is proportional to its variance. The covariance matrix is $\Sigma$, the lower bound of the truncation region is set to zero (enforcing non-negativity), and there is no upper bound (an infinite upper bound). The resulting matrix is normalized column-wise so that each column  (representing a signature) sums to one.

The exposures matrix and mutation counts matrix are simulated in the same manner as was done in the previous simulation study. That is:
$$E_{ij} \sim \text{Exponential}(0.001)$$

for all $i \in \{1, \dots, N\}, j \in \{1, \dots, G\}$ and 

$$M_{kg} \sim \text{Poisson}(\lambda_{kg} = (PE)_{kg})$$

for all $k \in \{1, \dots, 96\}, g \in \{1, \dots, G\}$.

\begin{figure}[htbp]
  \centering
  \begin{subfigure}[b]{0.45\textwidth}
      \centering
      \includegraphics[width=\textwidth]{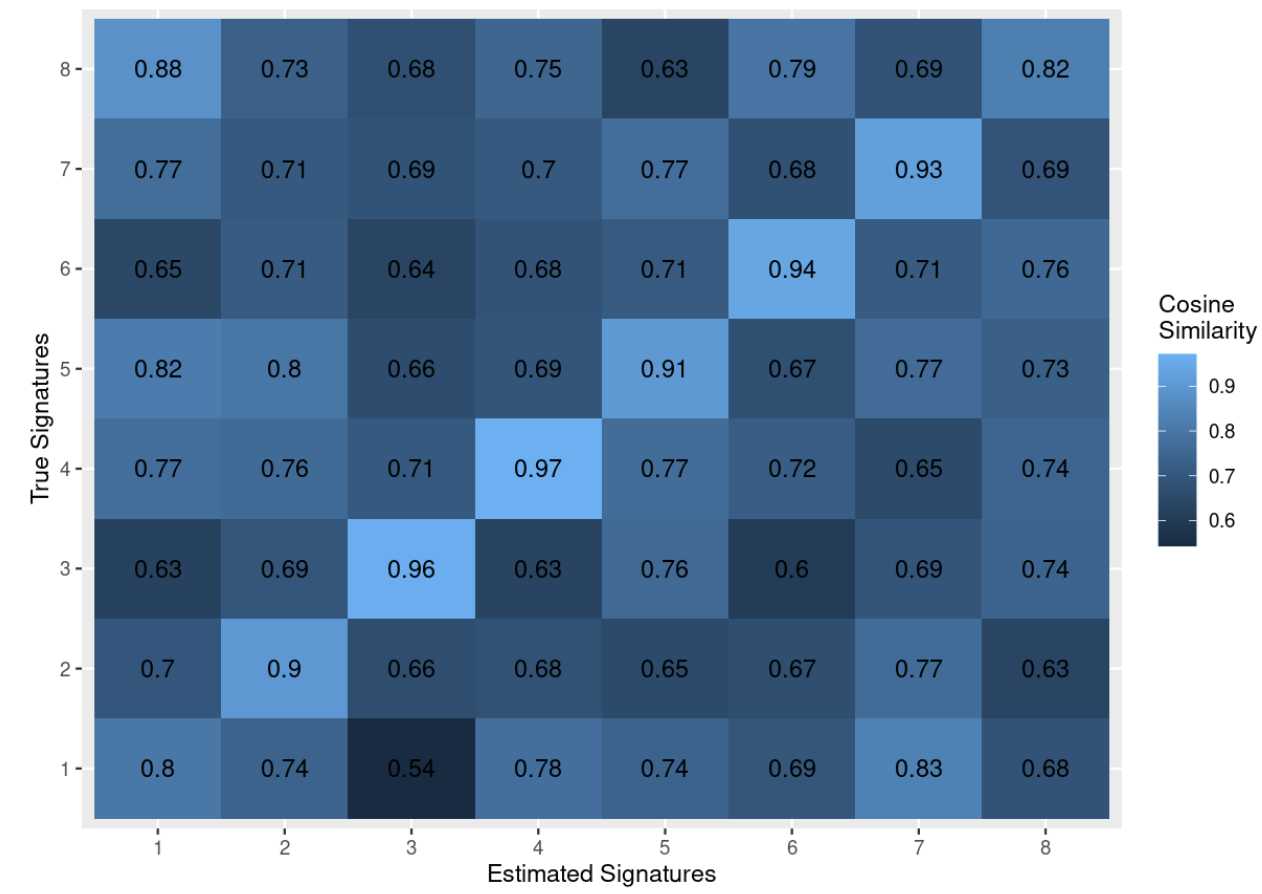}
      \caption{N=8 G=10 dataset \#1, MVN}
      \label{fig:seed223_mvn}
  \end{subfigure}
  \hfill
  \begin{subfigure}[b]{0.45\textwidth}
      \centering
      \includegraphics[width=\textwidth]{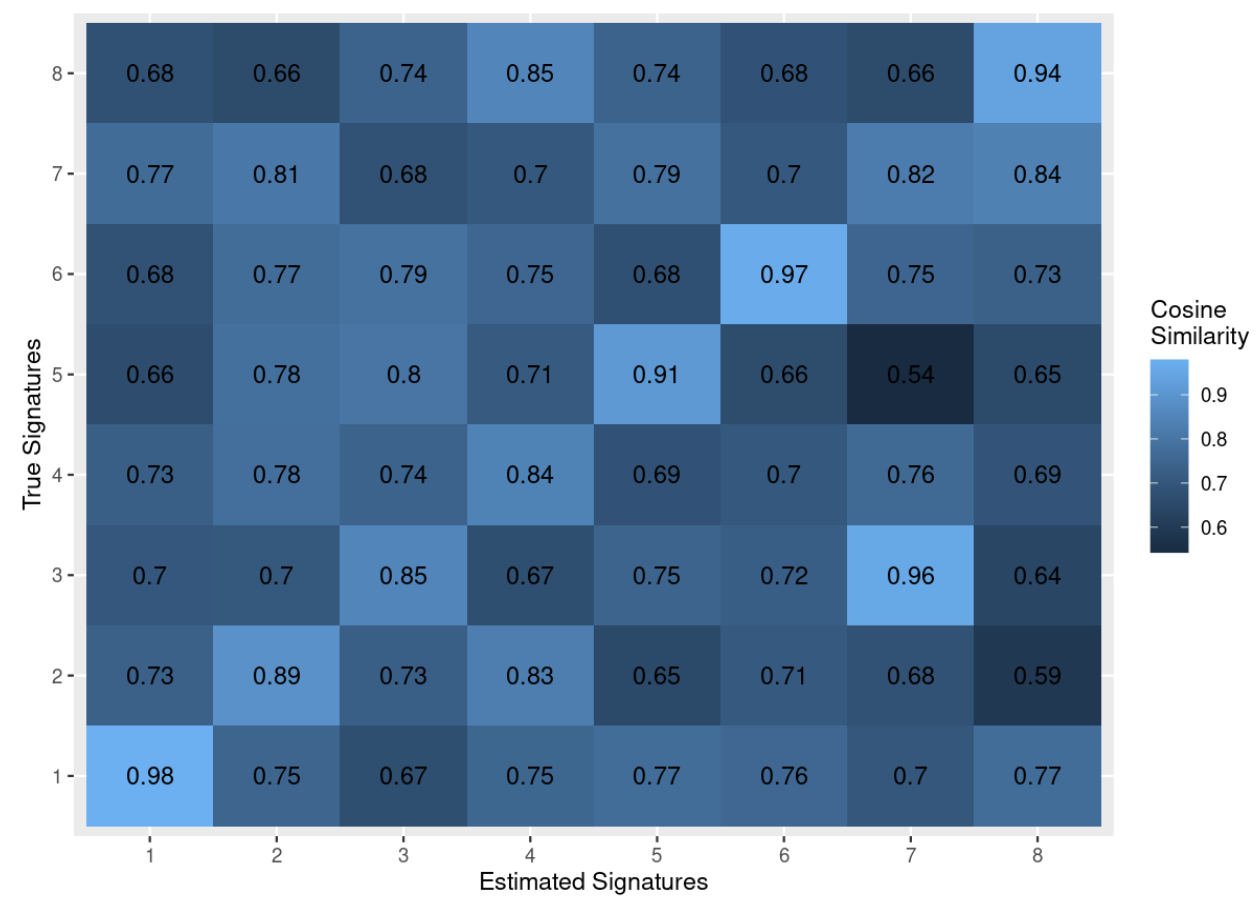}
      \caption{N=8 G=10 dataset \#1, nonMVN}
      \label{fig:seed223_nonmvn}
  \end{subfigure}
  \hfill
  \begin{subfigure}[b]{0.45\textwidth}
      \centering
      \includegraphics[width=\textwidth]{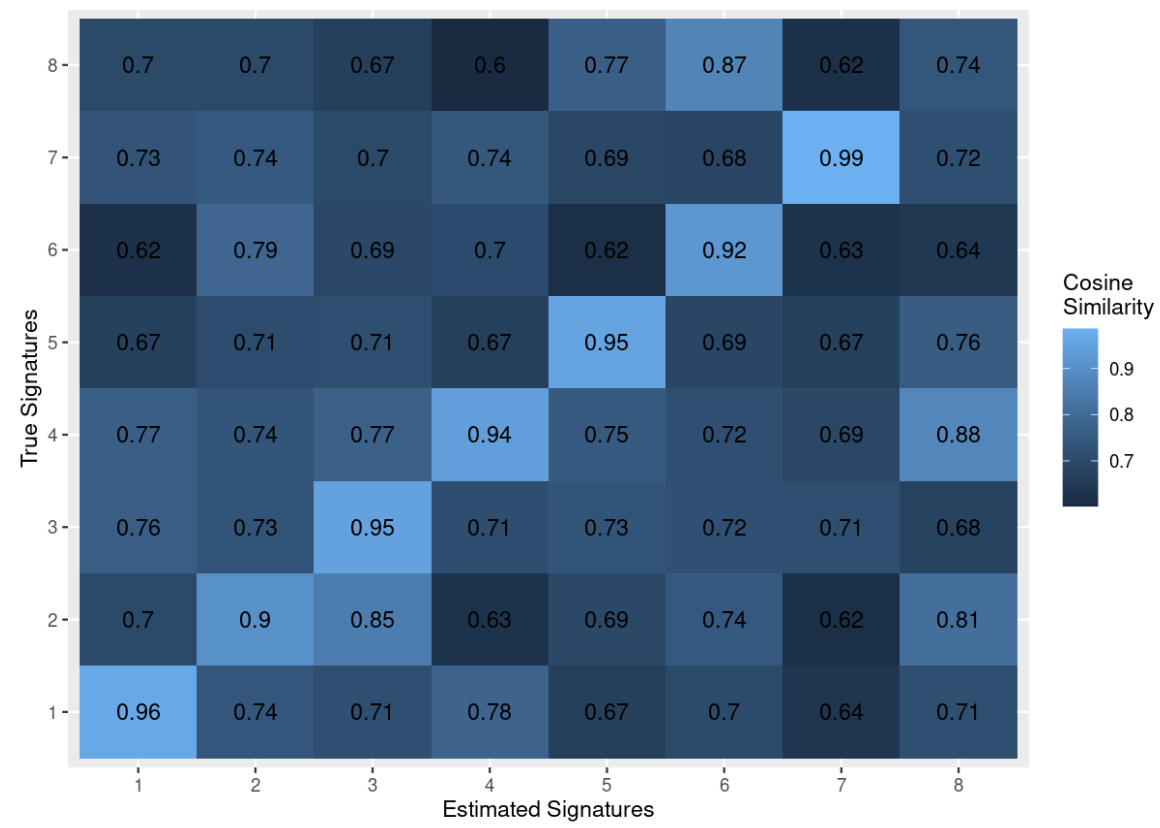}
      \caption{N=8 G=10 dataset \#2, MVN}
      \label{fig:seed423_mvn}
  \end{subfigure}
  \hfill
  \begin{subfigure}[b]{0.45\textwidth}
      \centering
      \includegraphics[width=\textwidth]{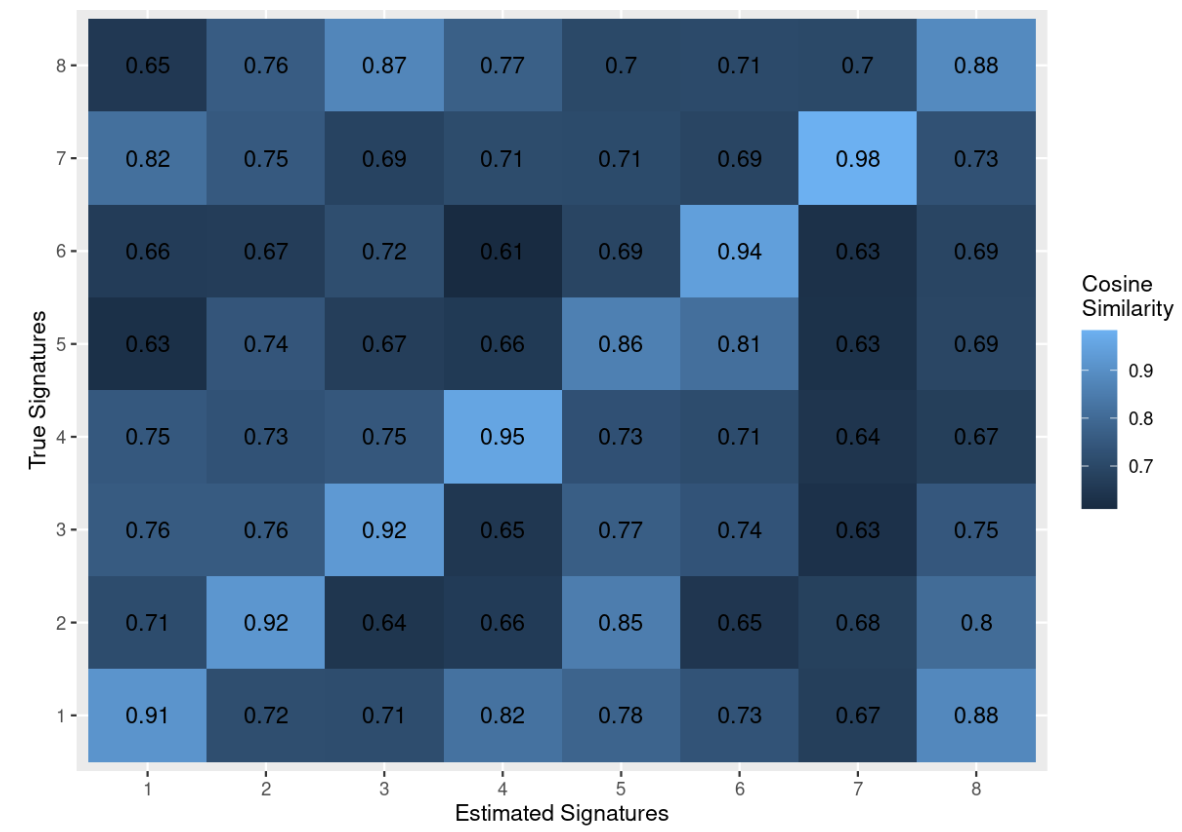}
      \caption{N=8 G=10 dataset \#2, nonMVN}
      \label{fig:seed423_nonmvn}
  \end{subfigure}
  \caption{Cosine similarity heatmap between estimated (x-axis) and COSMIC reference (y-axis) mutational signatures for simulation study of two distinct datasets each with sample size of 10 and rank (number of signatures) of 8. The datasets are simulated from the 3-parameter model, and the MVN and nonMVN methods are used. We expect a bright diagonal from the bottom left to top right corner in the heatmaps.}
  \label{fig:seed223_423_mvn_accuracy}
\end{figure}

We simulate two datasets in this manner and compare the performance of the MVN and nonMVN methods on each of the datasets, specifying the true rank of $N=8$. In terms of the accuracy of the estimated signatures, MVN consistently achieved comparable or better cosine similarities and RMSE values. In Figure~\ref{fig:seed223_423_mvn_accuracy}, we display the cosine similarity heatmaps for two of the datasets where the MVN method outperforms the nonMVN. We focus on the eight elements along the diagonal, using a cutoff of 0.9 to determine whether a signature has been correctly ``discovered''.

In the first dataset, it looks as though MVN has discovered six out of eight of the signatures while nonMVN has discovered four out of eight. In the second dataset, only one cosine similarity is below 0.9 for MVN while nonMVN has two below. Thus, this simulation study illustrates that indeed, there are instances in which the MVN method is preferable over nonMVN due to improved accuracy. Given that this dataset was constructed under the assumption of the 3-parameter hyperprior structure for the covariance matrix of the $P$ matrix, this leads us right into our Hierarchical model. Perhaps the Bayesian Hierarchical model will be able to uncover the parameters that generated this data.

\subsection{Hierarchical Model Simulation Study}

In this section, we apply our Bayesian Hierarchical methodology to a dataset where N=3 and G=60. The $P$, $E$, and $M$ matrices are all simulated using the same procedure discussed in the previous section. That is, the covariance matrix of the $P$ matrix is built from the parameterization where $\sigma_P^2=7, \rho_{\text{same}}=0.5, \rho_{\text{diff}}=-0.1$. Unlike the previous section, however, we have a smaller rank and a larger sample size. This is due to the fact that the hierarchical model is a much more computationally intensive model, taking several hours to run as opposed to a few seconds or minutes for the MVN or nonMVN methods.

\begin{figure}
\centering
\includegraphics[width=0.6\textwidth]{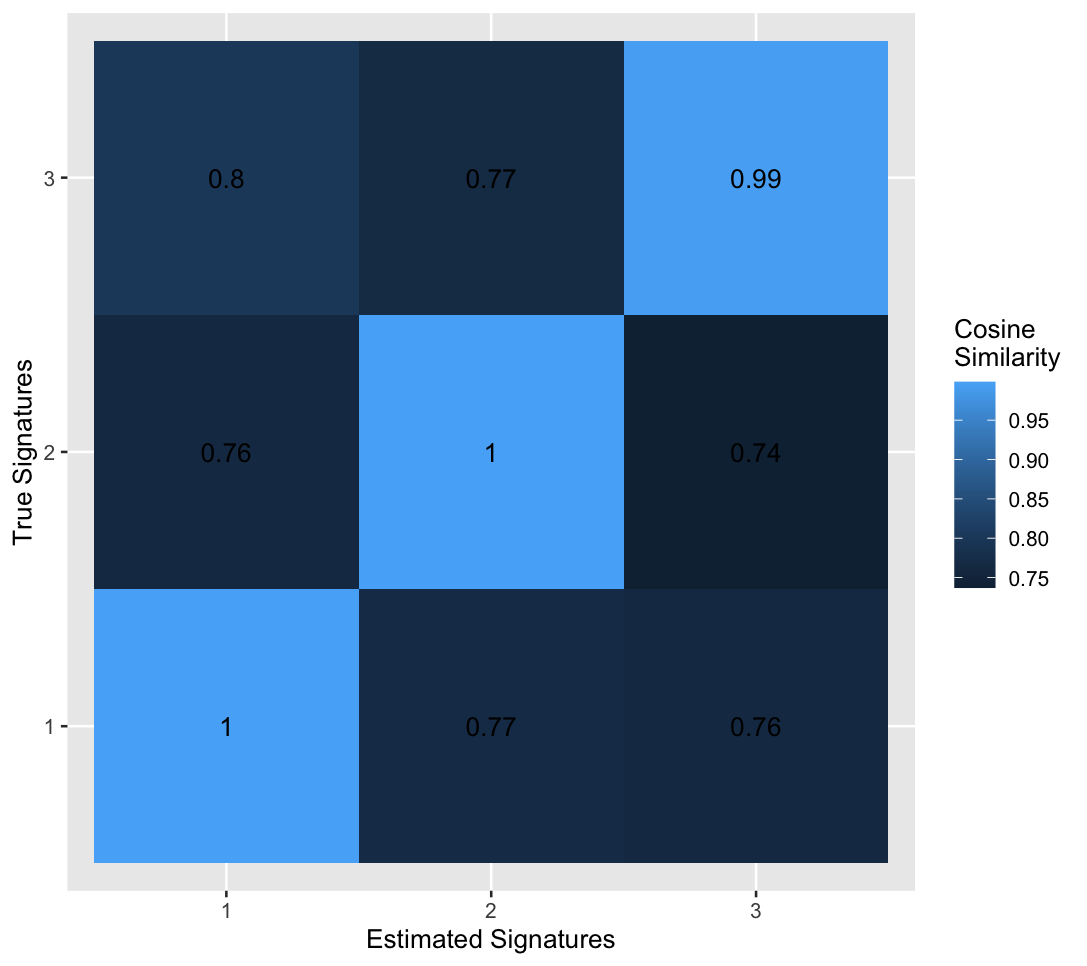}
\caption{Cosine similarity heatmap between estimated (x-axis) and COSMIC reference (y-axis) mutational signatures for simulation study with sample size of 120 and rank (number of signatures) of 3 using the Hierarchical method. We expect a bright diagonal from the bottom left to top right corner.}
\label{fig:n3g120_hier_cossim_heatmap}
\end{figure}

\begin{figure}[htbp]
    \centering
    \begin{subfigure}[b]{0.6\textwidth}
        \centering
        \includegraphics[width=\textwidth]{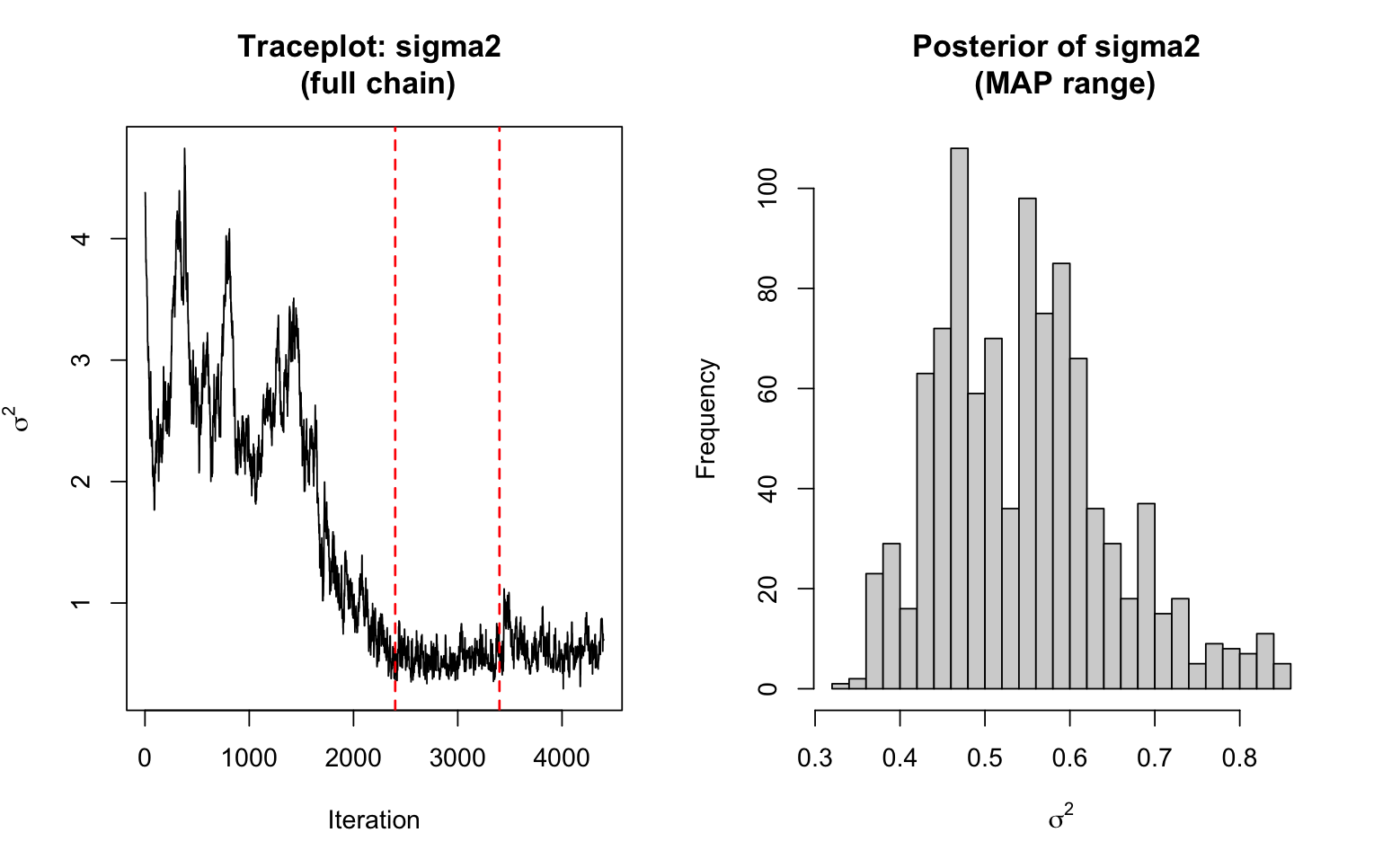}
    \end{subfigure}
    \begin{subfigure}[b]{0.6\textwidth}
        \centering
        \includegraphics[width=\textwidth]{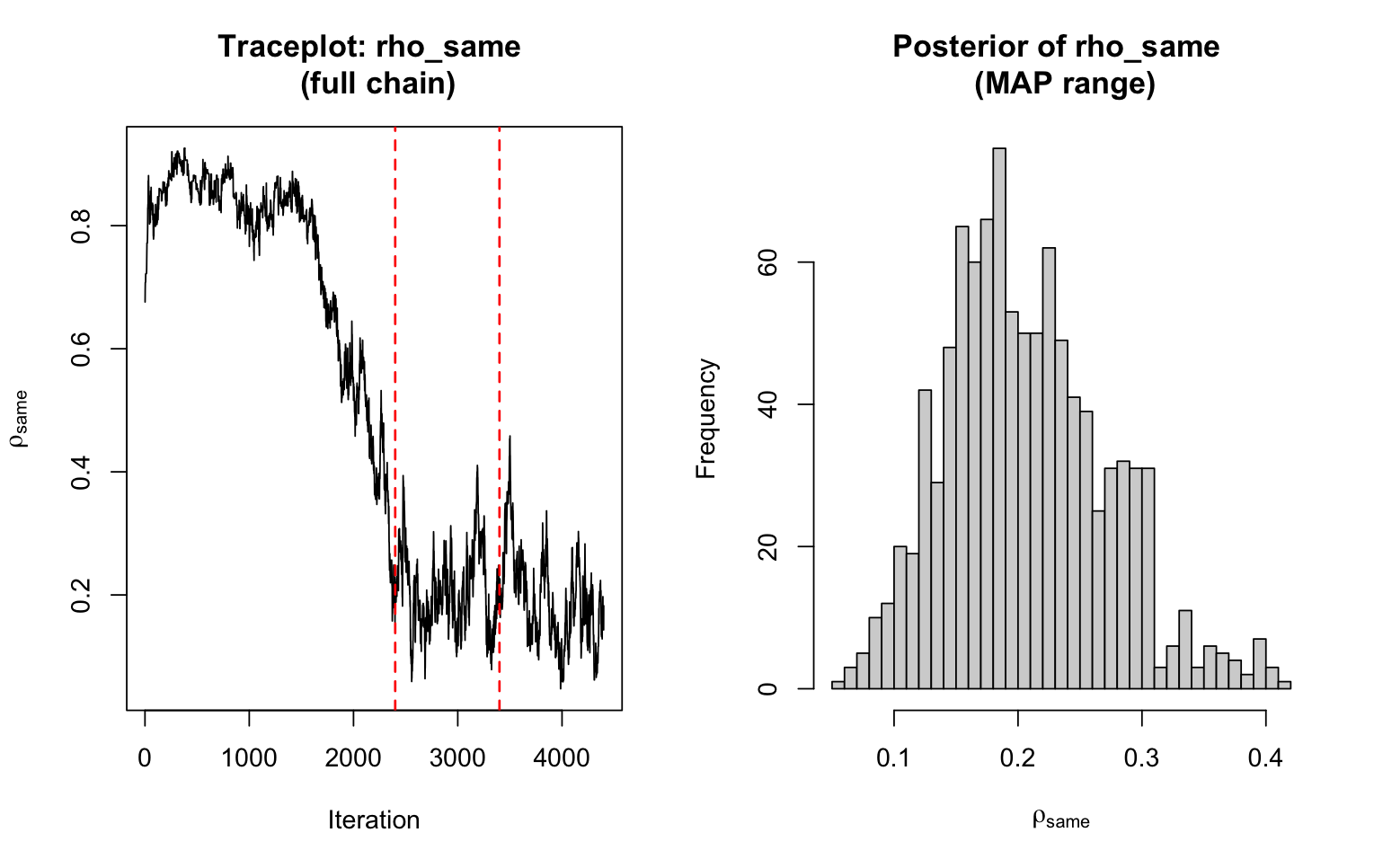}
    \end{subfigure}
    \hfill
    \begin{subfigure}[b]{0.6\textwidth}
        \centering
        \includegraphics[width=\textwidth]{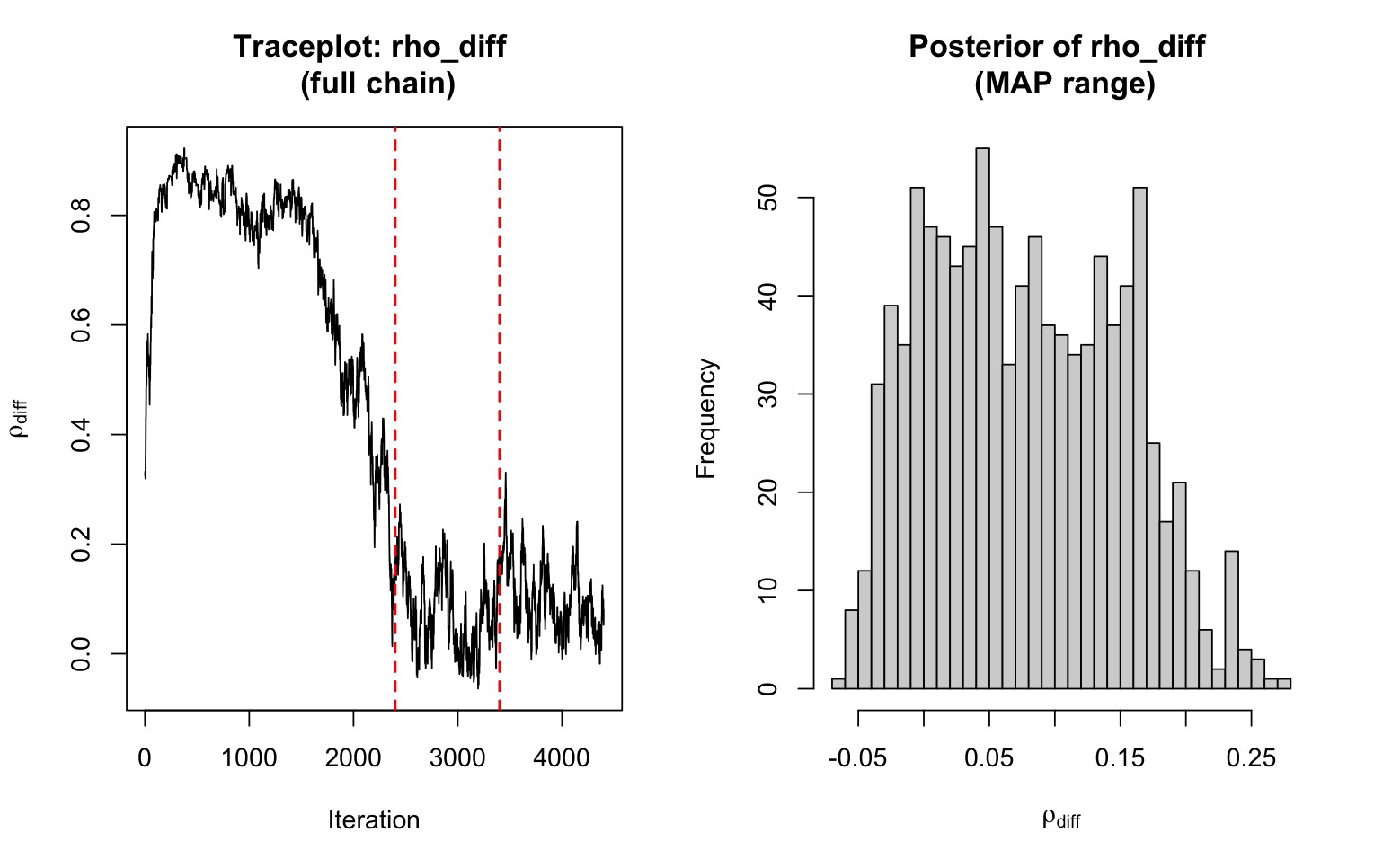}
    \end{subfigure}
    \caption{\textbf{Simulation Study} 
    Left figure: trace plots of full chain. Right figure: posterior distribution of the 1000 samples immediately before convergence for $\sigma_P^2, \rho_{\text{same}}, \rho_{\text{diff}}$ using Hierarchical model for dataset simulated from the 3-parameter structure ($\sigma_P^2=7, \rho_{\text{same}}=0.5, \rho_{\text{diff}}=-0.1$). The final 1000 iterations immediately preceding convergence are contained within the vertical, red, dotted lines in the trace plot figures.}
    \label{fig:n3g120_hier_traceplots}
\end{figure}

This model converged in 3400 iterations. The cosine similarity heatmap (Figure~\ref{fig:n3g120_hier_cossim_heatmap}) shows that the MAP estimate of the signatures matrix reconstructs the best aligned COSMIC reference signature almost perfectly, with all three cosine similarity values above 0.99. Figure~\ref{fig:n3g120_hier_traceplots} displays the trace plots for the three parameters of interest: the complete MCMC chains and the distributions of the final 1000 iterations immediately preceding convergence, which make up the MAP posterior samples.

Looking at the figure, we see the MCMC has done a good job exploring the parameter space. We see that the posterior distribution of $\rho_{\text{same}}$ seems to be either bimodal or unimodal with a slow convergence. The other two parameters', $\rho_{\text{diff}}$ and $\sigma_P^2$, distributions are right-skewed and converge more nicely. The maximum a-posteriori (MAP) estimates for each of the three parameters are:
$$\hat{\rho}_{\text{same}}=0.21, \; \hat{\rho}_{\text{diff}}=0.08, \; \hat{\sigma}_P^2 = 0.55$$

which result in $\hat{\rho}_{\text{same}}, \hat{\rho}_{\text{diff}}$ close to the original parameter values we specified ($0.5$ and $-0.1$, respectively) when simulating the covariance matrix of $P$. With a smaller sample size (of 30 or 60 instead of 120), the posterior mean of $\rho_{\text{same}}$ and $\rho_{\text{diff}}$ were further away from $0.5$ and $-0.1$, respectively, so indeed, a larger sample size is beneficial. The discrepancy between the simulation $\sigma_P^2 = 7$ and the estimated $\hat{\sigma_P}^2$ can be explained by the scale non-identifiability inherent in NMF. In NMF, we are reconstructing $M \approx PE$, but the matrices $P$ and $E$ can be rescaled in proportion to each other such that their product $PE$ remains the same. For instance, $PE = (2P)(E/2)$. As a result, it is possible for $P$ and $E$ to be rescaled during the MCMC algorithm which consequently will affect the scale of $\sigma_P^2$. In contrast, correlation parameters $\rho_{\text{same}}, \rho_{\text{diff}}$ are scale invariant.

\subsection{PCAWG Simulation Study}

We now apply our methods to a histology group in the Pan-Cancer Analysis of Whole Genomes (PCAWG) data: Panc-Endocrine. We use the latent rank estimated by \citet{landy2025bayesnmf}--6 signatures for Panc-Endocrine. In this analysis, while we do not have any ground truth values for the three parameters of the Hierarchical model, nor a ground truth set of reference signatures to verify our estimated signatures against, we can still evaluate our approaches by comparing results across multiple methods. By using the COSMIC reference signatures as the reference set, we can assess the consistency in estimated signatures and relative performance of the models.

\subsubsection{Panc-Endocrine}

We use MVN, nonMVN, and the Hierarchical model to estimate the Panc-Endocrine signatures from the Panc-Endocrine mutational counts matrix which contains 77 samples ($96 \times 77$).

\begin{figure}[htbp]
    \centering
    \begin{subfigure}[b]{0.55\textwidth}
        \centering
        \includegraphics[width=\textwidth]{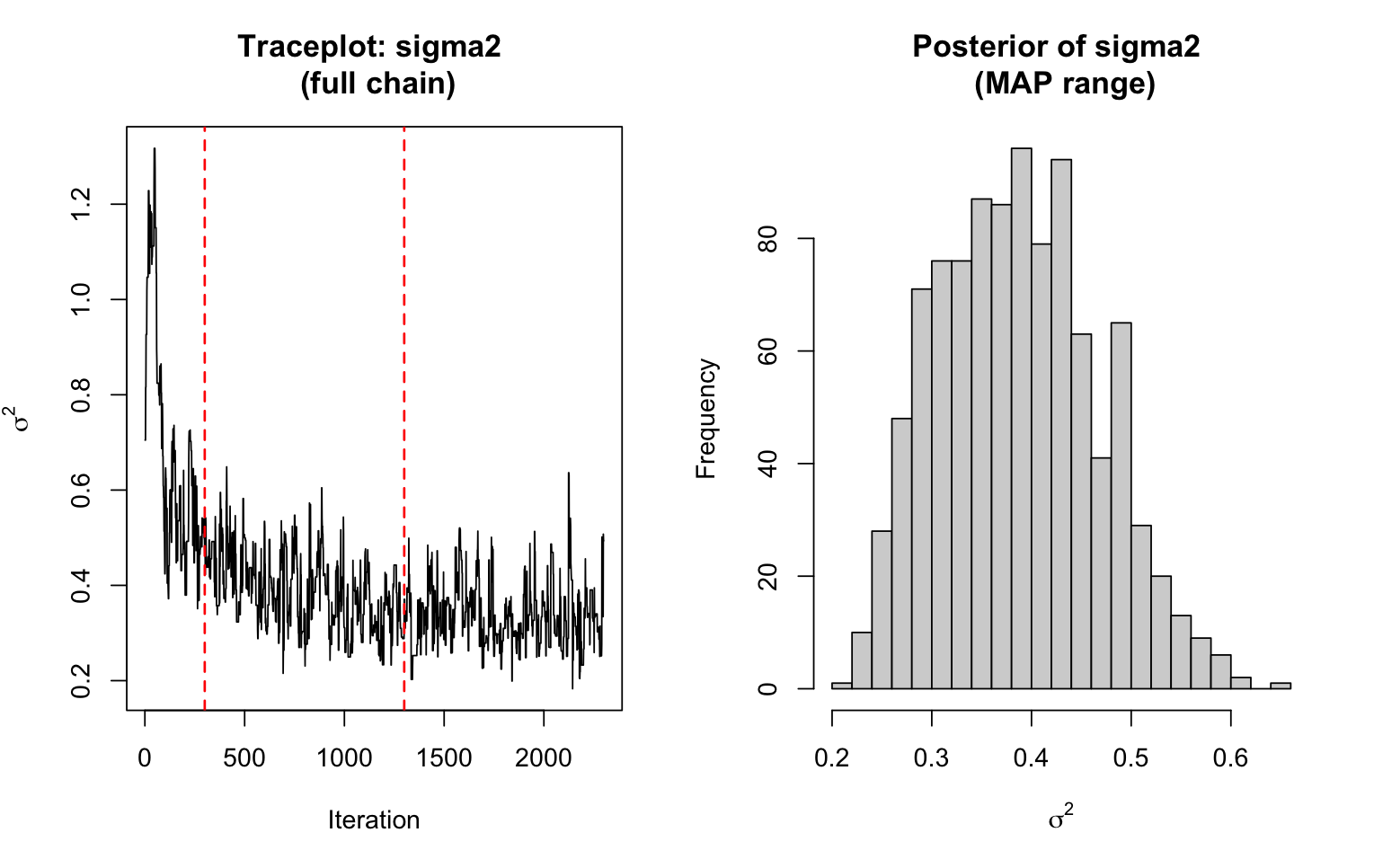}
    \end{subfigure}
    \begin{subfigure}[b]{0.55\textwidth}
        \centering
        \includegraphics[width=\textwidth]{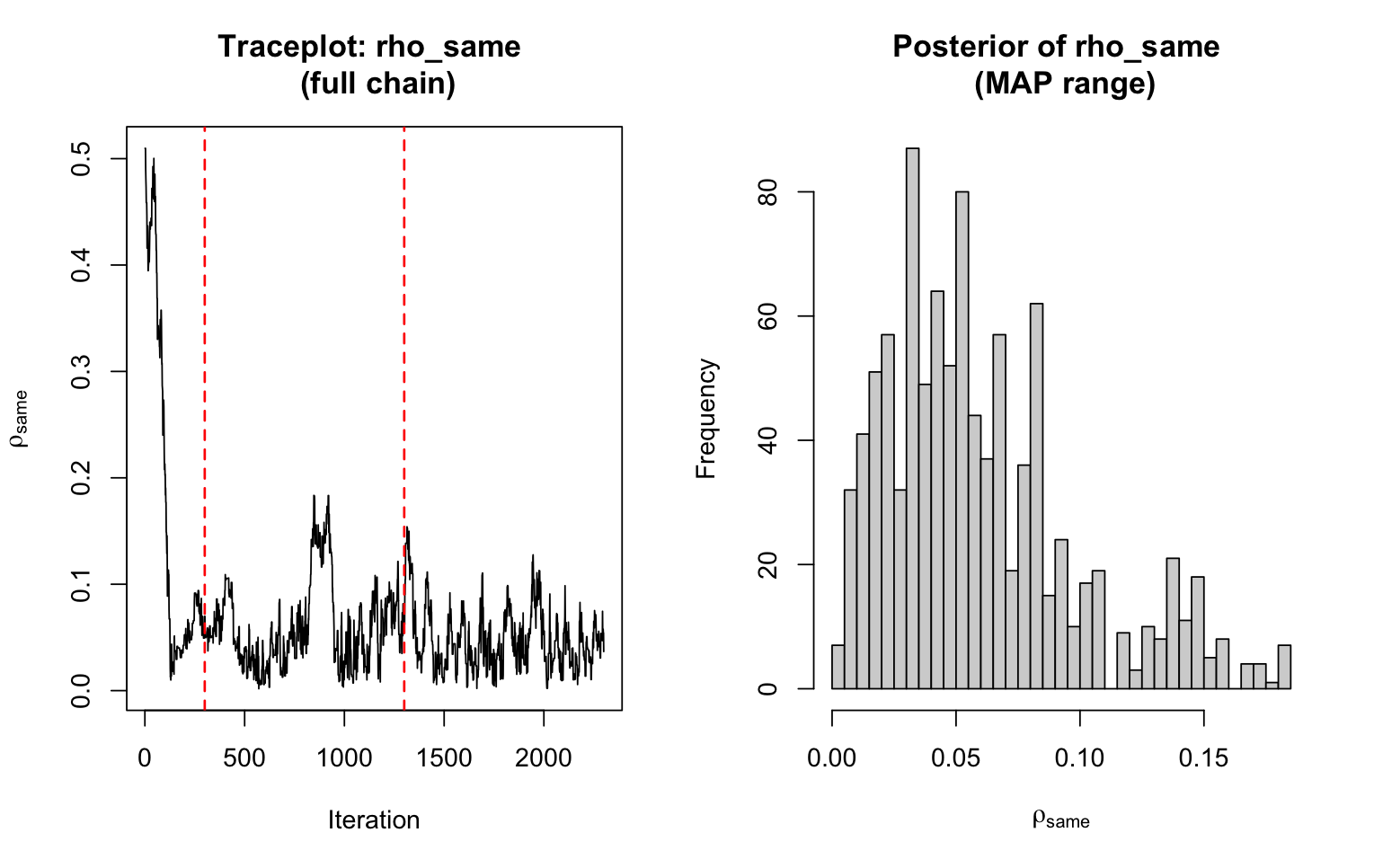}
    \end{subfigure}
    \hfill
    \begin{subfigure}[b]{0.55\textwidth}
        \centering
        \includegraphics[width=\textwidth]{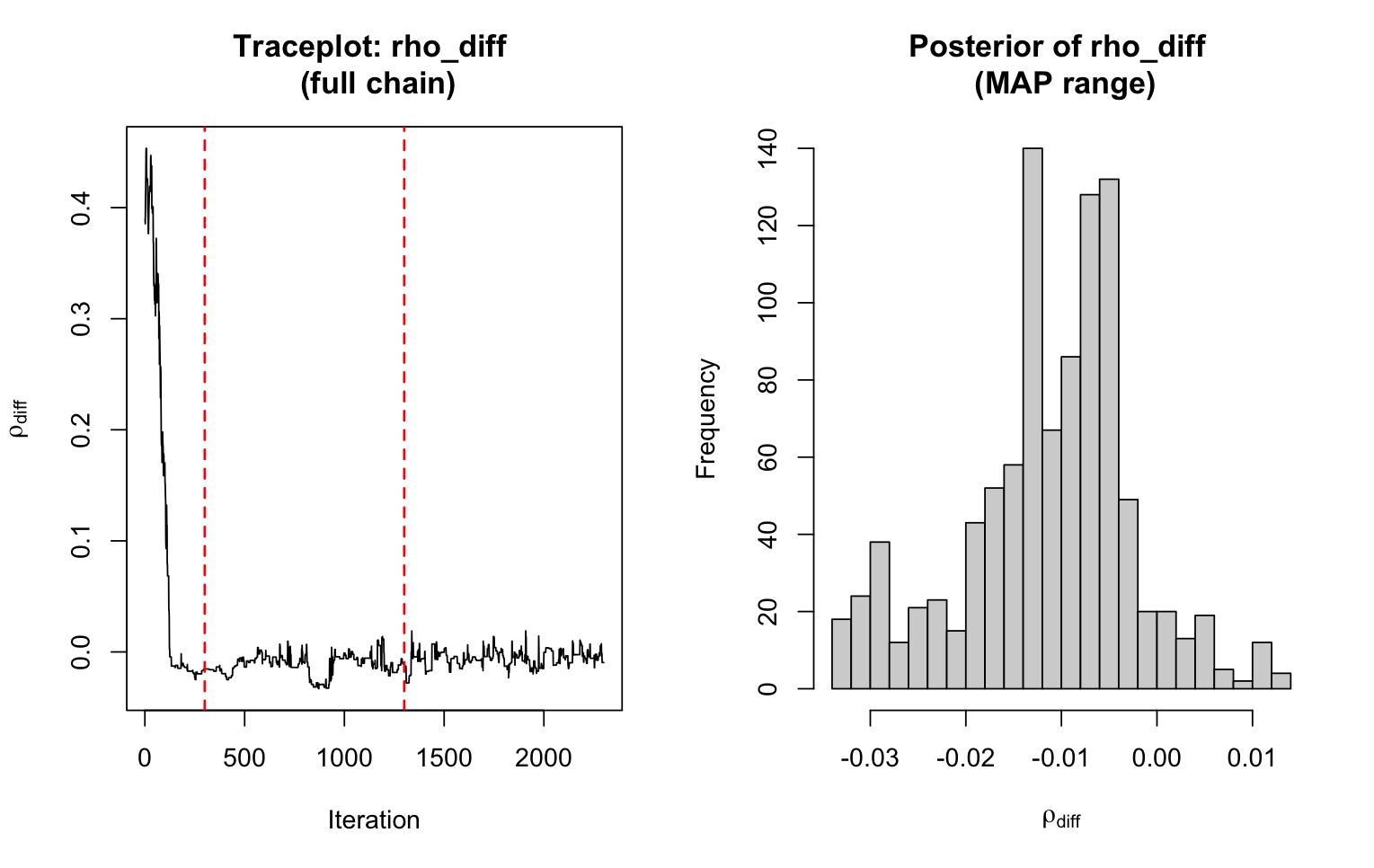}
    \end{subfigure}
    \caption{\textbf{Panc Endocrine data example}. Left figure: trace plots of full chain. Right figure: posterior distribution of the 1000 samples immediately before convergence for $\sigma_P^2, \rho_{\text{same}}, \rho_{\text{diff}}$ using Hierarchical model for Panc-Endocrine dataset. The final 1000 iterations immediately preceding convergence are contained within the vertical, red, dotted lines in the trace plot figures.}
    \label{fig:panc_endocrine_hier_traceplots}
\end{figure}

\begin{figure}[htbp]
  \centering
  \begin{subfigure}[b]{0.45\textwidth}
      \centering
      \includegraphics[width=\textwidth]{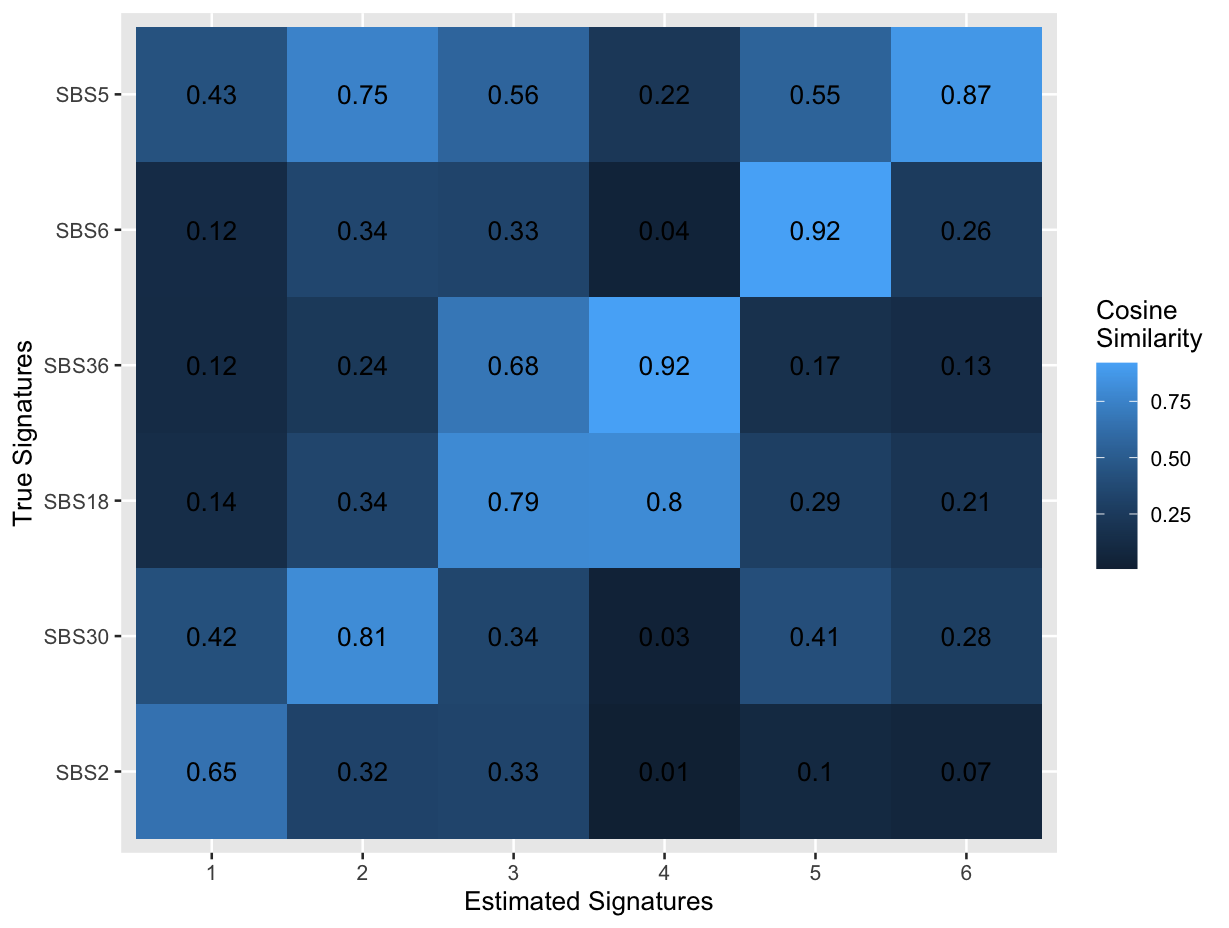}
      \caption{Panc-Endocrine, MVN}
      \label{fig:panc_endo_mvn_heatmap}
  \end{subfigure}
  \hfill
  \begin{subfigure}[b]{0.45\textwidth}
      \centering
      \includegraphics[width=\textwidth]{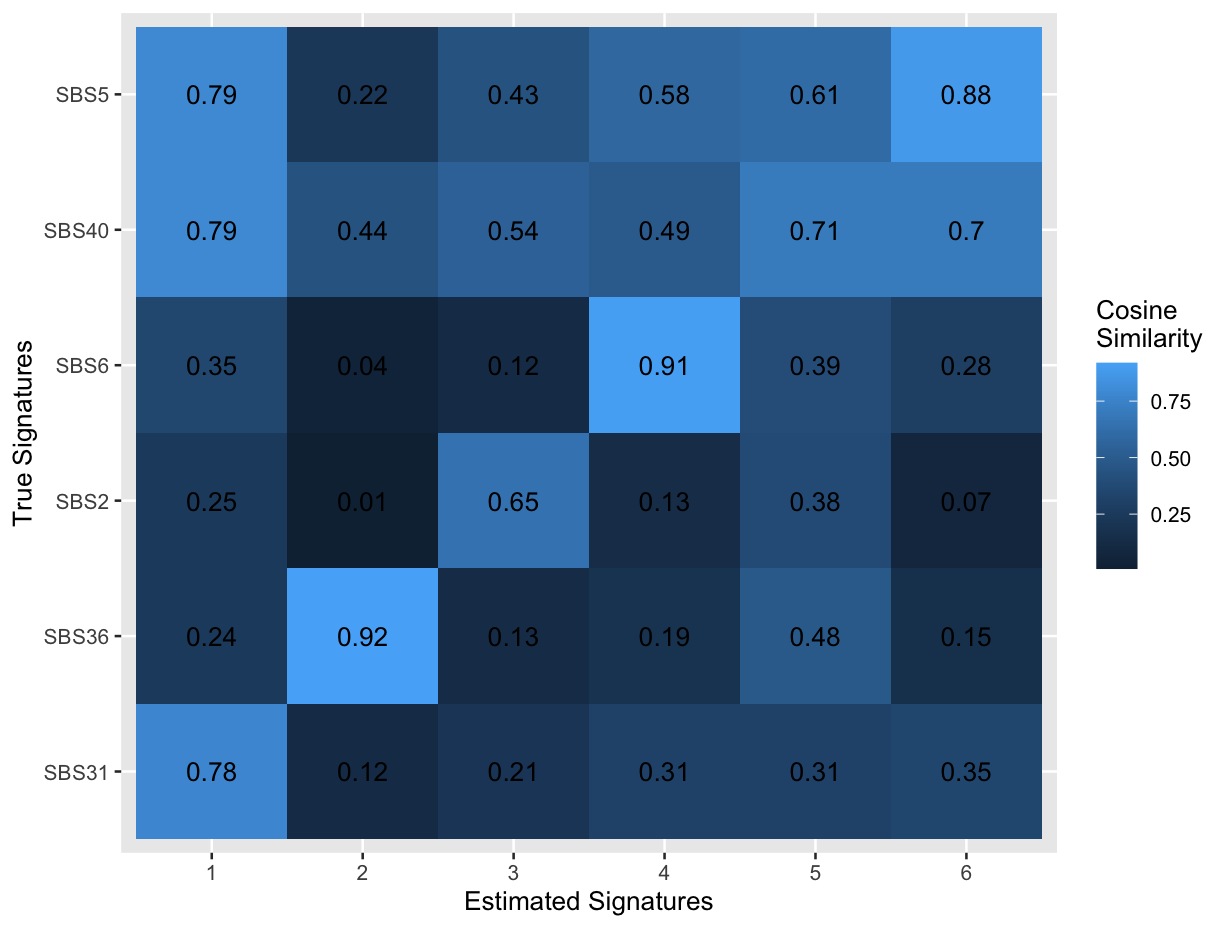}
      \caption{Panc-Endocrine, nonMVN}
      \label{fig:panc_endo_nonmvn_heatmap}
  \end{subfigure}
  \hfill
  \begin{subfigure}[b]{0.45\textwidth}
      \centering
      \includegraphics[width=\textwidth]{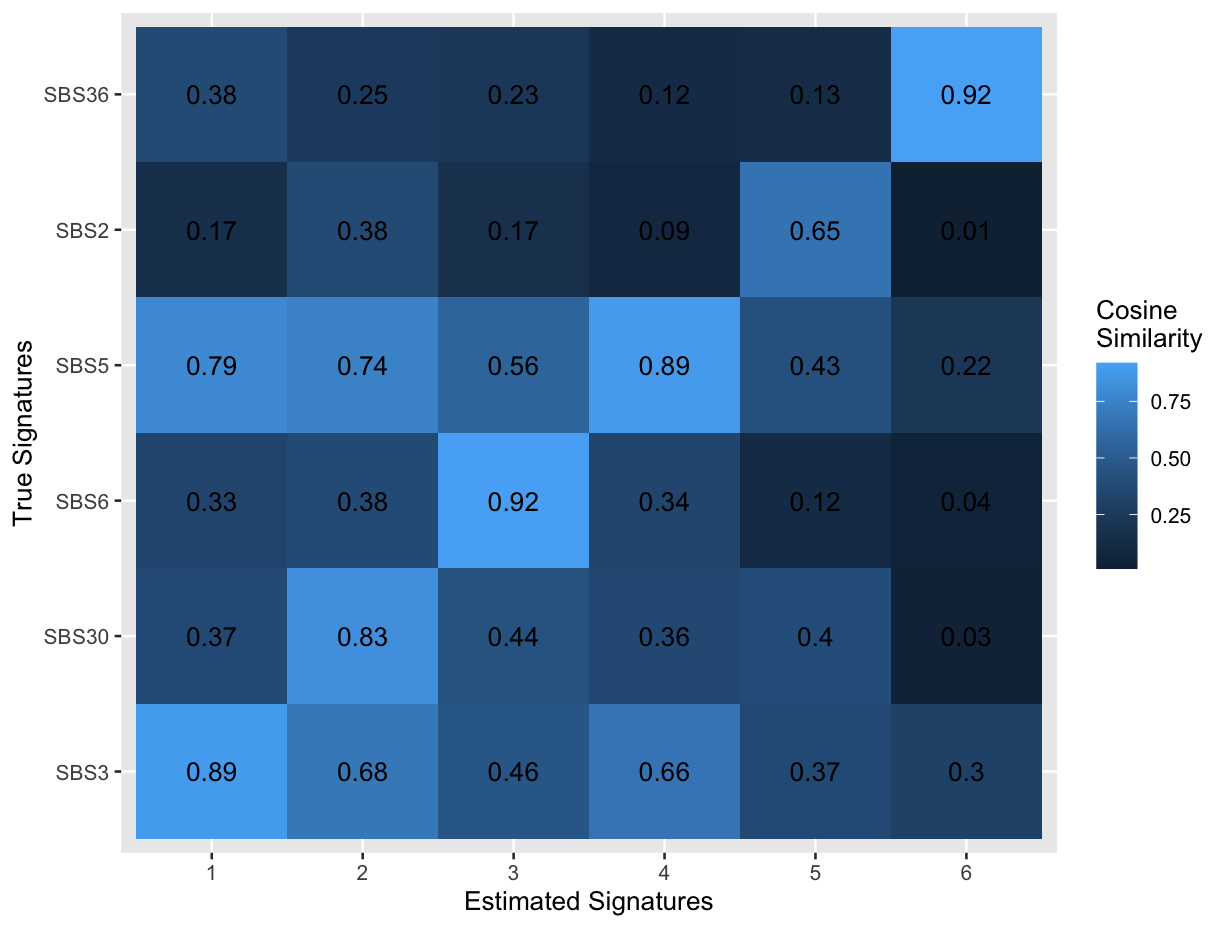}
      \caption{Panc-Endocrine, Hierarchical}
      \label{fig:panc_endo_hier_heatmap}
  \end{subfigure}
  \caption{\textbf{Panc Endocrine data example}. Cosine similarity heatmaps between estimated (x-axis) and COSMIC reference (y-axis) mutational signatures for Panc-Endocrine histology group. MVN, nonMVN, and Hierarchical methods with sample size of 77 and rank (number of signatures) of 6 are used. We expect a bright diagonal from the bottom left to top right corner.}
  \label{fig:panc_endo_all_heatmap}
\end{figure}

Looking at Figure~\ref{fig:panc_endocrine_hier_traceplots}, we see that the hierarchical model converges well for $\rho_{\text{same}}$ and $\rho_{\text{diff}}$, but surprisingly converges to values which are quite close to $0$ (MAP estimates $\hat{\sigma}_P^2=0.353, \hat{\rho}_{\text{same}}=0.053, \hat{\rho}_{\text{diff}}=-0.008$). However, if $\rho_{\text{same}}$ and $\rho_{\text{diff}}$ really were insignificant and approximately $0$, we would expect the performance of the hierarchical model to be similar to that of the nonMVN model (which assumes an identity matrix for the covariance of $P$). Looking at the heatmaps in Figure~\ref{fig:panc_endo_all_heatmap}, though, we see notable differences in the results across all three methods. 

Additionally, of note is that although the method does a slow job of learning (as around 2000 iterations are needed for convergence), there is evidence of our method being able to learn the true parameters, at least to a confined region. For example, most of the posterior density for $\rho_{\text{same}}$ lies between 0.0 and 0.1, indicating a subtle yet meaningful (nonzero) effect of same-center mutations. 

Lastly, the convergence control we use is defined to monitor the convergence of the log posterior. While this is likely to translate to the convergence of the $P$ and $E$ matrices as the biggest contributors to the posterior, it doesn't necessarily translate to the convergence of hyperprior parameters $\sigma_P^2$, $\rho_{\text{same}}$, and $\rho_{\text{diff}}$ since they contribute very little to the posterior. The trace plot for $\sigma_P$ in Figure~\ref{fig:panc_endocrine_hier_traceplots} suggests that $\sigma_P^2$ has not yet fully converged.

While the three models all identified signatures SBS3, SBS5, SBS6, and SBS36, there is variation in what the remaining two signatures are. Additionally, the Hierarchical model has five of six cosine similarity values at least as large as 0.83 while MVN and nonMVN both only have three of six that are that large. The distribution of the cosine scores along the diagonal of these heatmaps are not identical. So, even though the MAP estimates of the correlation parameters in the Hierarchical model may seem small in magnitude, Figure~\ref{fig:panc_endo_all_heatmap} illustrates strengths in signature estimation that the Hierarchical modeling framework can have when applied to the Panc-Endocrine dataset.

\section{Discussion}
\label{conclusion}

\subsection{Contributions}

In this thesis, we developed two novel Bayesian frameworks for mutational signature analysis that extend existing methods by explicitly modeling the dependence structure among mutation types. We extend the \texttt{bayesNMF} package to include both of these methods. We explored the Catalogue of Somatic Mutations in Cancer (COSMIC) signatures matrix to show both biological and statistical evidence, through simulated null distributions, of the existence of the dependence structure. We then formulate two Multivariate Truncated Normal models, one with fixed covariance from literature signatures and a Hierarchical Model that learns the dependence structure from the data. Finally, through simulation studies and application to PCAWG cancer types, we verify the accuracy of our methods and highlight strengths and weaknesses of the methods.

By modeling the signatures matrix as a Multivariate Truncated Normal, our approach shares much of the existing algorithmic framework with the Normal-Truncated Normal model, but it accounts for additional correlation structure that is omitted in existing models. Our MVN approach is particularly useful for smaller datasets, specifically, datasets where the samples-to-signatures ratio is small. Because there is less data, having a model that more accurately reflects the true model by incorporating a dependence structure between mutation types in the signatures matrix, rather than assuming independence, can be crucial. Indeed, in our simulation study with $8$ signatures and a sample size of $10$, the MVN signatures that were estimated had higher cosine similarities, on aggregate, compared to those of the nonMVN method. Additionally, a key strength of the MVN approach is the decrease in the number of iterations until the method converges when compared to the nonMVN method, using the same convergence control. Because the MVN method is more reflective of the true structure, it requires fewer iterations to attain the same Maximum A Posteriori (MAP) value; however, each iteration itself takes more time to compute due to the complexity introduced by the Truncated Multivariate Normal sampling.

The Hierarchical model, on the other hand, requires more data in order to correctly and efficiently learn the three parameters $\sigma_P^2, \rho_{same}, \rho_{diff}$. While there is no ground truth for the estimation of the three parameters in the PCAWG data application, we were able to verify the ability of our algorithm to uncover the true models by generating data from our specified structure. The Hierarchical model is also more computationally burdensome, but a shift from our current random-walk algorithm to a more sophisticated technique, such as adaptive Metropolis Hastings, may significantly lighten the burden.

\subsection{Limitations}

\subsubsection{Computational Challenges}

While the Truncated Multivariate Normal distribution is a conjugate distribution, theoretically solving for the full conditional does not translate so easily to the implementation of the method. In practice, efficiently exploring such a highly constrained parameter space and sampling from the Truncated MVN full conditional presents significant computational challenges. Each iteration of the MVN method can take around ten times as long as each iteration of the nonMVN method due to the complexity of sampling from the Truncated MVN full conditional distribution of the $P$ matrix which we implemented via Gibbs sampling in the \texttt{tmvtnorm} package \citep{R-tmvtnorm2023}. Although the current runtime burden does not preclude the effective use of our MVN method in simpler experimental settings, scaling the approach to applications with a larger number of signatures or an expanded mutation alphabet will significantly amplify the computational costs, making runtime efficiency a limitation.

Another drawback of our MVN method is our use of truncation bounds $\texttt{lower\_stricter}$ and $\texttt{upper\_stricter}$ for the Truncated MVN sampling which are set to 10 standard deviations above and below the mean. While these bounds help avoid numerical instability and ensure valid sampling by preventing the proposal of extreme values, they may not be optimal for all datasets or applications. Future work can focus on developing  a more robust, data-adaptive strategy for setting these bounds, or even eliminating them altogether through alternative sampling methods.

Finally, the main drawback of our Hierarchical model is the computational cost and runtime. Each iteration of the MCMC algorithm takes a couple of seconds, and with a few thousand iterations required for convergence (2000 for the PCAWG Panc-Endocrine simulation study, for instance), the overall runtime can extend to several hours on a single machine. As such, future work should focus on more efficient sampling methods to reduce per-iteration cost and overall runtime.

\subsection{Future Directions}

One immediate extension of our approaches is to expand the mutation alphabet to include additional flanking contexts. Our MVN method can be easily adapted by changing the dimensions of the signatures covariance matrix, but our Hierarchical model would require an explicit redefinition of its correlation structure to accommodate a larger set of mutation types. Including more flanking contexts could provide a more detailed view of the mutational processes at work as the model would have the ability to distinguish between processes that leave similar trinucleotide patterns but differ in extended sequence context.

As the volume of publicly available cancer datasets continues to expand, computational efficiency is becoming more critical. Our Hierarchical method currently uses a Metropolis-Hastings random walk with a fixed proposal standard deviation. Alternatively, in future works, leveraging a more sophisticated method such as Adaptive Rejection Metropolis Sampling (ARMS) (via the \texttt{armspp} R package) can be beneficial with regards to addressing the drawbacks of computational cost \citep{armspp1995}. ARMS dynamically adjusts the proposal distribution based on the local curvature of the log-posterior, eliminating the need to manually tune the proposal. This may improve overall sampling efficiency of our Hierarchical model which would result in the model being more robust and scalable for high-dimensional mutational signature analysis.

Similarly, advancements in sampling efficiency for the Truncated MVN distribution would improve the runtime of our MVN method. Although our approach converges in fewer iterations compared to the nonMVN method, the per-iteration runtime is more expensive. If we can optimize the per-iteration efficiency through improved sampling strategies, the overall MVN algorithm may ultimately achieve a faster runtime than the nonMVN algorithm through leveraging its rapid convergence while mitigating the cost per iteration.

Further refinements of the theoretical modeling for our Hierarchical model can be pursued in the future. One potential improvement is a more sophisticated parameterization that allows for different values of $\rho$ for each of the six mutation clusters suggested by our exploratory analyses, rather than a single $\rho_{\text{same}}$ across all mutation types. More ambitious alternatives, as described in earlier sections, such as using an Inverse Wishart or LKJ prior for the full signatures correlation matrix may offer greater flexibility and better capture the heterogeneity in correlation patterns.

Lastly, our current work has been applied to single-study data; however, extending our methods to a multi-study framework is a possible future direction. By incorporating multi-study NMF frameworks as laid out in Grabski et al. (2020) \citep{grabski2020bayesiancombinatorialmultistudyfactor}, we can jointly analyze data from multiple sources, which would increase statistical power and improve the generalizability of our mutational signature findings across different cancer cohorts.

\section{Acknowledgments}
The authors gratefully acknowledge that this work was supported by the NIH-NCI grant 5R01 CA262710-03, and NSF-DMS grant 2113707.

\newpage
\bibliographystyle{unsrtnat}
\bibliography{references}  






\newpage
\section{Supplementary Materials}

\subsection{Convergence Control}

Convergence of a Gibbs sampling algorithm can be defined in numerous ways, ranging from monitoring the stabilization of parameter estimates and log-likelihood values to more formal diagnostics such as the Gelman–Rubin statistic or effective sample size measures. In our approach, we implement a convergence control based on the stabilization of the maximum a posteriori (MAP) estimates. We define a convergence control (utilizing the \texttt{new\_convergence\_control} functionality in \texttt{bayesNMF}). This implements a heuristic based on tracking the maximum a posteriori (MAP) value found by the sampler. We let the maximum number of iterations be \texttt{maxiters}=10000, \texttt{MAP\_over}=1000, and \texttt{MAP\_every}=100 (by default) \citep{landy2025bayesnmf}.

During the MCMC run, every \texttt{MAP\_every} iterations we compute the log posterior (or an unnormalized posterior density) of the MAP estimate over the previous \texttt{MAP\_over} iterations and update a running maximum if this sample has a higher posterior probability than any seen so far. If we observe that no change in MAP posterior probability value has been found (the chain has stayed within 0.1\% of the previous value), we interpret this as a sign that the chain has likely converged to the high-probability region and is exploring around the optimum. In other words, the sampler has stabilized in terms of finding better solutions. At that point, we can terminate the sampling.

The parameters \texttt{MAP\_every} and \texttt{MAP\_over} are set based on a balance between sensitivity and stability: \texttt{MAP\_every} should not be too large (so that we check often enough), and \texttt{MAP\_over} should be large relative to the autocorrelation time so that we give the chain ample opportunity to find a new mode if one exists. In our experiments, values like \texttt{MAP\_every} = 100 and \texttt{MAP\_over} = 500–1000 worked well.

\subsection{Implementation with Truncated MVN Sampler: Tuning}

While the theoretical derivation of the posterior distributions for our MVN model follows nicely from conjugacy, practical implementation requires careful tuning of the truncated multivariate normal sampler. We use the \texttt{rtmvnorm} function from the \texttt{tmvtnorm} package (2023) to sample from the posterior of the $P$ matrix \citep{R-tmvtnorm2023}. This function offers several parameters that we adjust to ensure robust sampling, particularly in high-dimensional spaces where extreme values or non-positive definite covariance matrices can be problematic.

\paragraph{Algorithm Choice:}
We set the \texttt{algorithm} parameter to ``gibbs". The Gibbs sampling algorithm is particularly well-suited for high-dimensional truncated distributions as it samples each component conditionally. The package also provides an option to use rejection sampling; however, in high-dimensional spaces, rejection sampling can be very inefficient with low acceptance rates.

\paragraph{Precision Matrix:} Rather than supplying the covariance matrix directly, we provide its inverse (the precision matrix). Using the precision matrix leads to improved numerical stability and computational efficiency. In particular, Gibbs sampler formulas are much simpler in terms of the precision matrix than in terms of the covariance matrix \citep{R-tmvtnorm2023}. It is necessary, however, that the covariance matrix is positive definite, so in our implementation, we ensure that the covariance matrix is made positive definite (by adding a small jitter to its diagonal if necessary) before inverting it.

\paragraph{Start Value:} By default, if a start.value is not specified, the Gibbs sampler initializes each component at the nearest finite bound (or zero if a bound is infinite). Even if zero is a valid point in the domain, it may be an extreme corner relative to the distribution’s center. If the mean vector has any moderately large positive entries, starting at $0$ forces the first update to “leap” into the far upper tail to reach the mean. As such, we set start.value so that the initial state is near the mean (clamped within the allowed support) which avoids beginning at $0$ if $0$ is far in the tail of the prior.

\paragraph{Burn-In Samples:}
The parameter \texttt{burn.in.samples} is set to $10$. This means that the first $10$ samples generated by the Gibbs sampler are discarded, allowing the chain to move away from its initial state and reach a region that better represents the target distribution.

\paragraph{Lower and Upper Truncation Bounds:}
Instead of using the default lower and upper bounds of zero and $\infty$ for the truncation region, we restrict the sampling region to a range that is 10 standard deviations above and below the mean vector. This modification adapts the bounds to the scale of each signature, as determined by the standard deviation from the covariance matrix of $P$, and significantly enhanced numerical stability for our simulation studies and experiments. When bounds are set too far from the mean relative to the scale, the probability of obtaining a valid sample can diminish to the point of numerical instability, potentially leading to NaN or infinite values during sampling.

\subsection{Normal-Truncated Normal Single-study Full Conditional Derivation}
\label{Appendix: Norm-Trunc Norm Posterior Derivation}

We have $K$ mutation types, $G$ tumor samples, and $N$ signatures. The mutation counts matrix: $M \in \mathbb{R}_{\geq 0}^{K \times G}$, the signatures matrix: $P \in \mathbb{R}_{\geq 0}^{K \times N}$, the exposures matrix: $E \in \mathbb{R}_{\geq 0}^{N \times G}$.

We have:
\begin{align*}
    P_{kn} &\sim \text{TruncNorm}(\mu_{n}^P, \sigma_{kn}^{2P}, 0, \infty) \\
    E_{ng} &\sim \text{TruncNorm}(\mu_{ng}^E, \sigma_{ng}^{2E}, 0, \infty) \\
    M_{kg} &\sim \text{N}((PE)_{kg}, \sigma_k^2)
\end{align*}
$$\sigma_k^2 \sim \text{InvGamma}(\alpha_k, \beta_k)$$

First, we derive the full conditional for $\sigma_k^2$:
\begin{align*}
    p(\sigma_k^2 | P, E, M) &\propto p(\sigma_k^2) \prod_k \prod_g M_{kg} \\
    &\propto \frac{{\beta_k}^{\alpha_k}}{\Gamma(\alpha_k)}(\sigma_k^2)^{-\alpha_k-1} \exp\left\{-\frac{\beta_k}{\sigma_k^2}\right\} \cdot \frac{1}{(\sigma_k \sqrt{2\pi})^G} \cdot \exp\left\{-\frac{1}{2\sigma_k^2} \sum_g (M_{kg}-(PE)_{kg})^2\right\} \\
    &\propto (\sigma_k^2)^{-\alpha_k - G/2 - 1} \exp\left\{-\frac{1}{\sigma_k^2}(\beta_k + \frac{1}{2} \sum_g (M_{kg} - (PE)_{kg})^2)\right\} \\
    &\sim \text{InvGamma}\left(\alpha_k + \frac{G}{2}, \beta_k + \frac{1}{2} \sum_g (M_{kg} - (PE)_{kg})^2\right)
\end{align*}

Now, we derive the full conditional for $E$:
\begin{align*}
    p(E_{ng} | E_{-ng}, P, M, \sigma_k) &\propto p(E_{ng})\prod_k \prod_g M_{kg} \\
    &\propto \mathbb{1} \{E_{ng} \geq 0\} \exp\left\{-\frac{1}{2\sigma_{ng}^{2E}} (E_{ng}-\mu_{ng}^E)^2\right\} \cdot \exp\left\{-\sum_k \frac{1}{2\sigma_k^2} (M_{kg}-(PE)_{kg})^2\right\} \\
    &\propto \mathbb{1} \{E_{ng} \geq 0\} \exp\left\{-\frac{1}{2\sigma_{ng}^{2E}} (E_{ng}-\mu_{ng}^E)^2 -\sum_k \frac{1}{2\sigma_k^2} (P_{kn}E_{ng} - (M_{kg}-\hat{M}_{kg(-n)}))^2\right\} \\
\end{align*}
$$\propto \mathbb{1}\{E_{ng} \geq 0\}\exp\left\{E_{ng}^2 \left(-\frac{1}{2\sigma_{ng}^{2E}} - \frac{1}{2\sigma_k^2} \sum_k P_{kn}^2\right) + E_{ng}\left(2\left(\frac{1}{2\sigma_{ng}^{2E}}\right)(\mu_{ng}^E) + 2\sum_k \left(\frac{1}{2\sigma_k^2}\right) P_{kn}(M_{kg}-\hat{M}_{kg(-n)})\right)\right\}$$
$$\propto \mathbb{1}\{E_{ng} \geq 0 \} \exp\left\{-\frac{1}{2}\left(\frac{1}{\sigma_{ng}^{2E}} + \sum_k \frac{p_{kn}^2}{\sigma_k^2}\right) \left(E_{ng} - \frac{\mu_{ng}^E/\sigma_{ng}^{2E} + \sum_k \frac{P_{kn}}{\sigma_k^2} (M_{kg} - \hat{M}_{kg(-n)})}{1/\sigma_{ng}^{2E} + \sum_k \frac{P_{kn}^2}{\sigma_k^2}}\right)^2\right\}$$
$$\sim \text{TruncNorm}\left(\frac{\mu_{ng}^E/\sigma_{ng}^{2E} + \sum_k \frac{P_{kn}}{\sigma_k^2} (M_{kg} - \hat{M}_{kg(-n)})}{1/\sigma_{ng}^{2E} + \sum_k \frac{P_{kn}^2}{\sigma_k^2}}, \frac{1}{\sigma_{ng}^{2E}} + \sum_k \frac{p_{kn}^2}{\sigma_k^2}, 0, \infty\right)$$

Finally, we derive the full conditional for $P$:
\begin{align*}
    p(P_{kn} | P_{-kn},E,M,\sigma_k) &\propto p(P_{kn})\prod_k \prod_g p(M_{kg} | \cdots) \\
    &\propto p(P_{kn})\prod_g p(M_{kg} | \cdots) \\
    &\propto \mathbb{1}\{P_{kn}\geq0\}\exp\left\{\frac{1}{2\sigma_{kn}^{2P}}(P_{kn}-\mu_{kn}^P)^2\right\}\exp\left\{-\frac{1}{2\sigma_k^2}\sum_g (M_{kg}-(PE)_{kg})^2\right\}
\end{align*}
$$\propto \mathbb{1}\{P_{kn}\geq0\} \exp\left\{P_{kn}^2 \left(-\frac{1}{2\sigma_k^{2P}} - \frac{1}{2\sigma_k^2}\sum_g E_{ng}^2\right) + P_{kn} \left(2\left(\frac{1}{2\sigma_{kn}^{2P}}\right)(\mu_{kn}^P) + 2\left(\frac{1}{2\sigma_k^2}\right) \sum_g E_{ng} (M_{kg} - \hat{M}_{kg(-n)})\right)\right\}$$
$$\propto \mathbb{1}\{P_{kn}\geq0\} \exp\left\{-\frac{1}{2} \left(\frac{1}{\sigma_{kn}^{2P}}+\frac{1}{\sigma_k^2}\sum_g E_{ng}^2\right)\left(P_{kn} - \frac{\mu_{kn}^P / 2\sigma_{kn}^{2P} + (\sum_g E_{ng} (M_{kg} - \hat{M}_{kg(-n)})) / \sigma_k^2}{\frac{1}{\sigma_{kn}^{2P}}+\frac{1}{\sigma_k^2}\sum_g E_{ng}^2}\right)^2\right\}$$
$$\sim \text{TruncNorm}\left( \frac{\mu_{kn}^P / 2\sigma_{kn}^{2P} + (\sum_g E_{ng} (M_{kg} - \hat{M}_{kg(-n)})) / \sigma_k^2}{\frac{1}{\sigma_{kn}^{2P}}+\frac{1}{\sigma_k^2}\sum_g E_{ng}^2}, \frac{1}{\frac{1}{\sigma_{kn}^{2P}}+\frac{1}{\sigma_k^2}\sum_g E_{ng}^2}, 0, \infty\right)$$

\subsection{Normal-Truncated MVN Single-study Full Conditional Derivation}
\label{Appendix: MVN-Truncated MVN derivation}

We have $K$ mutation types, $G$ tumor samples, and $N$ signatures. The mutation counts matrix: $M \in \mathbb{R}_{\geq 0}^{K \times G}$, the signatures matrix: $P \in \mathbb{R}_{\geq 0}^{K \times N}$, the exposures matrix: $E \in \mathbb{R}_{\geq 0}^{N \times G}$.

The $P$ matrix now has an additional parameter, its covariance matrix across mutation types: $\Sigma \in \mathbb{R}^{K \times K}$, which is the main focus of this project. 

We can rewrite:
$$M_{kg} \sim N((PE)_{kg}, \sigma_k^2)$$
$$\sigma_k^2 \sim \text{InvGamma}(\alpha_k, \beta_k)$$

in terms of a MVN distribution. Define $\Sigma^M \in \mathbb{R}^{K \times K}$ (the covariance matrix, across mutation types, of $M^T$) for $j, k \in \{1, \dots, 96\}$:
\begin{align*}
    \Sigma_{jk}^M &= \begin{cases}
    \sigma_k^2 & \forall j = k  \\
    0 & \forall j \neq k 
\end{cases} \\
 &= \begin{bmatrix}
   \sigma_1^2 & 0 & \dots  \\ 
   0 & \ddots & \\ 
   \vdots &  & \sigma_{96}^2
 \end{bmatrix}
\end{align*}

So, we have $\vv{p_n} \in \mathbb{R}^{K \times 1}$ (column of $P$, n-th signature), $\vv{M_g} \in \mathbb{R}^{K \times 1}$ (column of $M$, g-th tumor sample):
\begin{align*}
    \vv{p_n} &\sim \text{MVNTruncNorm}(\vv{\mu_n}^P, \Sigma, 0, \infty) \\
    E_{ng} &\sim \text{TruncNorm}(\mu_{ng}^E, \sigma_{ng}^{2E}, 0, \infty) \\
    \vv{M_g} &\sim \text{MVN}((\vv{PE})_g, \Sigma^M)
\end{align*}

We derive the full conditional form for the $P$ matrix. Note that the full conditional for $E$ and $\sigma_k^2$ remain the same as in the case where $P$ has no covariance matrix (see Appendix~\ref{Appendix: Norm-Trunc Norm Posterior Derivation}).

The likelihood of $M$ is:
\begin{align*}
    \prod_k \prod_g p(M_{kg} | \dots) &\propto \exp\left\{\sum_k \sum_g \left(-\frac{1}{2\sigma_k^2}\right)(M_{kg}-(PE)_{kg})^2\right\} \\
    &\propto \exp\left\{\sum_k \sum_g \left(-\frac{1}{2\sigma_k^2}\right)(P_{kn}E_{ng} - (M_{kg}-\hat{M}_{kg(-n)}))^2\right\} \\
    &\propto \exp\left\{\sum_{k=1}^{96} \left(-\frac{1}{2\sigma_k^2}\right)\left(P_{kn}^2 \sum_g E_{ng}^2 - 2P_{kn} \sum_g E_{ng} (M_{kg}-\hat{M}_{kg(-n)})\right)\right\}  \tag{where $\hat{M}_{kg(-n)} = \sum_i P_{ki}E_{ig} - P_{kn}E_{ng}$} \\
    &\propto \exp\left\{\sum_{k=1}^{96} \left(-\frac{1}{2\sigma_k^2}\right)\left(P_{kn}^2 \sum_g E_{ng}^2 - 2P_{kn} A_k^*\right)\right\} \tag{let $A_k^* = \sum_g E_{ng} (M_{kg}-\hat{M}_{kg(-n)})$} \\
    &= \exp\left\{-\frac{1}{2} \left[(\sum_g E_{ng}^2) \vv{p_n}^T (\Sigma^M)^{-1} \vv{p_n} - 2 \vv{p_n}^T (\Sigma^M)^{-1} \vv{A^*}\right] \right\}
\end{align*}

The prior of $\vv{p_n}$ is:
\begin{align*}
    p(\vv{p_n}) &\propto \exp\left\{-\frac{1}{2}(\vv{p_n}-\vv{\mu_n}^p)^T \Sigma^{-1} (\vv{p_n}-\vv{\mu_n}^p)\right\}
\end{align*}

Combining the two, we obtain the full conditional for $\vv{p_n}$:
\begin{align*}
    p(\vv{p_n} | P_{-\vv{n}}, E, M, \sigma_k) &\propto \exp\left\{-\frac{1}{2} \left[ (\vv{p_n}-\vv{\mu_n}^p)^T \Sigma^{-1} (\vv{p_n}-\vv{\mu_n}^p) + (\sum_g E_{ng}^2) \vv{p_n}^T (\Sigma^M)^{-1} \vv{p_n} - 2 \vv{p_n}^T (\Sigma^M)^{-1} \vv{A^*} \right]\right\} \\ 
    = &\exp\left\{-\frac{1}{2} \left[ \vv{p_n}^T\Sigma^{-1}\vv{p_n} - 2(\vv{p_n})^T\Sigma^{-1}\vv{\mu_n}^P + (\vv{\mu_n}^P)^T(\Sigma^{-1})(\vv{\mu_n}^P) + (\sum_g E_{ng}^2) \vv{p_n}^T (\Sigma^M)^{-1} \vv{p_n} - 2 \vv{p_n}^T (\Sigma^M)^{-1} \vv{A^*} \right] \right\} \\
    = &\exp\left\{-\frac{1}{2} \left[ \vv{p_n}^T\left(\Sigma^{-1} + (\sum_g E_{ng}^2)(\Sigma^M)^{-1}\right)\vv{p_n} - 2\vv{p_n}^T\left(\Sigma^{-1}\vv{\mu_n}^P + (\Sigma^M)^{-1} \vv{A^*}\right) + c \right] \right\} \tag{where $c$ is some constant wrt $\vv{p_n}$} \\
    = &\exp\left\{-\frac{1}{2} \left[ \vv{p_n}^T A \vv{p_n} - 2\vv{p_n}^T B + c \right] \right\} \tag{where $A = \Sigma^{-1} + (\sum_g E_{ng}^2)(\Sigma^M)^{-1}$ and $B = \Sigma^{-1}\vv{\mu_n}^P + (\Sigma^M)^{-1} \vv{A^*}$} \\
    = &\exp\left\{-\frac{1}{2} \left[ \vv{p_n}^T A \vv{p_n} - 2\vv{p_n}^T B + B^T(A^{-1})^T B - B^T (A^{-1})^T B \right] \right\} \\
    = &\exp\left\{-\frac{1}{2} \left[ (\vv{p_n} - A^{-1}B)^T A (\vv{p_n} - A^{-1}B ) + c' \right] \right\}
\end{align*}

This implies that the full conditional of $\vv{p_n}$ is distributed as $\text{MVNTruncNorm}(\vv{\mu_n}^{new}, \Sigma^{new}, 0, \infty)$ where:
\begin{align*}
    \vv{\mu_n}^{new} &= A^{-1}B = \left(\Sigma^{-1} + (\sum_g E_{ng}^2)(\Sigma^M)^{-1}\right)^{-1} \left(\Sigma^{-1}\vv{\mu_n}^P + (\Sigma^M)^{-1} \vv{A^*}\right) \\
    \Sigma^{new} &= A^{-1} = \left(\Sigma^{-1} + (\sum_g E_{ng}^2)(\Sigma^M)^{-1}\right)^{-1}
\end{align*}

Note that this matches the derivation from Appendix~\ref{Appendix: Norm-Trunc Norm Posterior Derivation} for $P$ if $\Sigma$ is a diagonal matrix.

The full conditional derivations for $E$ and $\sigma_k^2$ are the same as in Appendix~\ref{Appendix: Norm-Trunc Norm Posterior Derivation}.

\end{document}